\documentclass[11pt]{article}

\usepackage{cite}
\usepackage{amssymb,amsmath}
\usepackage{latexsym}
\usepackage{mathrsfs}
\usepackage{hyperref} 
\usepackage{verbatim}
\usepackage{graphicx}

\def\x{{\bf x}}
\def\y{{\bf y}}
\def\z{{\bf z}}
\def\k{{\bf k}}
\def\p{{\bf p}}

\def\eps{\epsilon}

\def\R{{\mathcal R}}
\def\P{{\mathcal P}}

\def\bea{\begin{eqnarray}}
\def\eea{\end{eqnarray}}
\def\be{\begin{equation}}
\def\ee{\end{equation}}
\def\ba{\begin{array}}
\def\ea{\end{array}}
\def\nn{\nonumber}

\begin{document}

\setlength\arraycolsep{2pt}

\renewcommand{\theequation}{\arabic{section}.\arabic{equation}}
\setcounter{page}{1}

\begin{titlepage}

\begin{center}

\vskip 1.0 cm

{\LARGE  \bf Consistency relations for sharp features in the primordial spectra}

\vskip 1.0cm

{\large
Sander Mooij${}^{a}$, Gonzalo A. Palma${}^{a}$, Grigoris Panotopoulos${}^{a}$ and Alex Soto${}^{b}$
}

\vskip 0.5cm

{\it
${}^{a}$Departamento de F\'{i}sica, Facultad de Ciencias F\'{i}sicas y Matem\'{a}ticas, Universidad de Chile\\ 
\mbox{Blanco Encalada 2008, Santiago, Chile} \\
\mbox{${}^{b}$Departamento de F\'{i}sica, Facultad de Ciencias, Universidad de Chile}\\ 
\mbox{Las Palmeras 3425, \~{N}u\~{n}oa, Santiago, Chile} \\
}

\vskip 1.5cm

\end{center}

\begin{abstract}

We study the generation of sharp features in the primordial spectra within the framework of effective field theory of inflation, wherein curvature perturbations are the consequence of the dynamics of a single scalar degree of freedom. We identify two sources in the generation of features: rapid variations of the sound speed $c_s$ (at which curvature fluctuations propagate) and rapid variations of the expansion rate $H$ during inflation. With this in mind, we propose a non-trivial relation linking these two quantities that allows us to study the generation of sharp features in realistic scenarios where features are the result of the simultaneous occurrence of these two sources. This relation depends on a single parameter with a value determined by the particular model (and its numerical input) responsible for the rapidly varying background. As a consequence, we find a one-parameter consistency relation between the shape and size of features in the bispectrum and features in the power spectrum. To substantiate this result, we discuss several examples of models for which this one-parameter relation (between $c_s$ and $H$) holds, including models in which features in the spectra are both \emph{sudden} and \emph{resonant}.

\end{abstract}

\end{titlepage}

\newpage

\section{Introduction} 
\setcounter{equation}{0}

Current Cosmic Microwave Background (CMB) and Large Scale Structure (LSS) observations favor cosmological initial conditions consistent with a Gaussian distribution of curvature perturbations, parametrized by an almost scale invariant power spectrum. In the absence of better explanations, these observations are usually considered to support the idea that our universe underwent an early stage of accelerated expansion~\cite{Guth:1980zm}, in the form of canonical single-field slow-roll inflation~\cite{Linde:1981mu, Albrecht:1982wi}, wherein primordial perturbations evolved adiabatically in a quasi-de Sitter background, driven by the evolution of a scalar field gently descending the slope of a smooth scalar potential. Such models generically predict an almost Gaussian distribution of primordial fluctuations with a slightly red tilted power spectrum~\cite{Starobinsky:1980te, Mukhanov:1981xt} in addition to a small value for the tensor to scalar ratio, in agreement with the latest CMB constraints~\cite{Ade:2013uln, Ade:2015lrj}.

It is possible, however, that primordial fluctuations had their origin under less placid conditions, with their dynamics characterized by time-scales shorter than the inverse of the quasi-de Sitter expansion rate $H$, breaking the simple statistical properties offered by single-field slow-roll scenarios. If present, such time-scales are expected to be associated to the existence of massive degrees of freedom (with masses larger than $H$) either affecting the inflationary background or interacting with curvature perturbations as they exit the horizon during inflation~\cite{Achucarro:2010da}. The landmark of models incorporating time-scales shorter than $H^{-1}$ is the emergence of features in the primordial spectra, with their size and shape cha\-rac\-te\-ri\-zing the physics associated to the high energy degrees of freedom that participated in their production~\cite{Adams:1997de, Bean:2008na, Cespedes:2012hu, Battefeld:2013xka, Adshead:2013zfa, Noumi:2013cfa, Mizuno:2014jja, Ashoorioon:2014yua}. Consequently, the existence of features in both the primordial power spectrum and bispectrum could represent a unique chance of acquiring a better understanding of inflation by giving us insights into mass scales ---additional to $H$--- characterizing the fundamental theory responsible for them~\cite{Chluba:2015bqa}. 

There are well known examples of inflationary models allowing for features in the primordial spectra. One obvious class corresponds to canonical single-field models in which the inflaton field traverses a sudden change in the slope of its potential, for example in the form of a step or a bump, resulting in a brief interruption of the slow-roll dynamics. This type of dynamics leads to a sudden variation of the expansion rate, without necessarily implying large deviations from the quasi-de Sitter background as the leading geometrical configuration during inflation. These sudden variations of the background are inevitably felt by the primordial fluctuations while they exit the horizon, leaving a set of features imprinted in the power spectrum and bispectrum~\cite{Starobinsky:1992ts, Adams:2001vc, Gong:2005jr, Covi:2006ci, Ashoorioon:2006wc, Ashoorioon:2008qr, Chialva:2008xh, Miranda:2013wxa, Romano:2014kla, Novaes:2015uza, Cai:2015xla,Jain:2008dw,Jain:2009pm}. The same mechanism leads to the production of features in $P(X,\phi)$ models if the $\phi$-dependent part of the Lagrangian has a feature that interrupts the slow-roll evolution of the field as it traverses the target space of the theory. In this case, not only the expansion rate will display a sudden variation in time, but also the sound speed $c_s$ at which curvature perturbations propagate (typically encountered in non-canonical realizations of inflation such as $P(X,\phi)$-models), leading to a richer variety of features imprinted in the primordial spectra as compared to those produced in canonical models of inflation.  

Another class of models allowing for features in the primordial spectra are multi-field models of inflation,\footnote{Yet another mechanism leading to the emergence of features ---also involving additional degrees of freedom--- is offered by particle production during inflation~\cite{Chung:1999ve, Elgaroy:2003hp, Mathews:2004vu, Romano:2008rr, Barnaby:2009dd, Fedderke:2014ura}. We will not be concerned about this type of models in this article.} in which the inflaton background trajectory meanders the landscape offered by the multi-field target space~\cite{Gordon:2000hv, GrootNibbelink:2000vx, GrootNibbelink:2001qt, Tye:2008ef, Tye:2009ff}, enhancing the interaction between curvature perturbations and fields orthogonal to the trajectory each time the trajectory undergoes a bend~\cite{Tolley:2009fg, Cremonini:2010sv, Cremonini:2010ua, Achucarro:2010da, Cespedes:2012hu, Battefeld:2014aea}. The resulting interactions have the chance of generating localized features in the spectra if the bend-rate is larger than the Hubble expansion rate $H$ during inflation~\cite{Gao:2015aba}. In this case there are two main regimes leading to features:

\begin{enumerate}
\item[I.] {\it Features generated by multiple degrees of freedom}: If the time variation of the bend-rate of the trajectory is comparable to ---or larger than--- the mass of the fields orthogonal to the trajectory, these become excited, generating features in the spectra with a wavelength characterized by the mass of the excited degrees of freedom~\cite{Shiu:2011qw, Cespedes:2012hu, Chen:2014cwa}. These degrees of freedoms may be heavy or light, depending on whether their masses are larger or smaller than  the inflationary expansion rate $H$.

\item[II.] {\it Features generated by a single degree of freedom}: On the other hand, if the time variation of the bend-rate is smaller than the mass of the fields orthogonal to the trajectory, then the only dynamically relevant degree of freedom consists of curvature perturbations~\cite{Achucarro:2012sm, Achucarro:2012yr}. It is then possible to study these features using an effective single-field theory describing the evolution of curvature perturbations, in which the background is parametrized by the sound speed $c_s$ of curvature perturbations and the expansion rate $H$~\cite{Achucarro:2010jv, Baumann:2011su, Achucarro:2012sm, Avgoustidis:2012yc, Achucarro:2012yr, Burgess:2012dz, Gwyn:2012mw, Cespedes:2013rda, Castillo:2013sfa, Gwyn:2014doa, Gong:2014rna}. In this case, the features in the primordial spectra are found to be the consequence of the time variation of $c_s$ and $H$ resulting from the bend in multi-field space~\cite{Achucarro:2010da, Cespedes:2012hu, Saito:2013aqa, Gao:2013ota}. 
\end{enumerate}
While in the second regime the background is of the multi-field type, the dynamics associated to the fluctuations maps into an effective single-field theory description. This is possible because the heavy degrees of freedom may decouple from the light curvature perturbations even if the bend rate is much larger than the expansion rate~\cite{Cespedes:2012hu}. Nonetheless, the multi-field dynamics associated to the heavy degrees of freedom continue to be present through the background, leading to the emergence of nontrivial couplings in the effective field theory for the fluctuations, such as the sound speed $c_s$ describing the propagation of curvature perturbations.

Despite of the complexity involved in the emergence of features, the study of models allowing for them may be systematically organized ---and substantially simplified--- by adopting the effective field theory (EFT) framework~\cite{Cheung:2007st, Senatore:2010wk} in which one disregards the details of the background dynamics to focus, instead, on the dynamics of curvature perturbations and any other degrees of freedom relevant around horizon crossing. To be more specific, the EFT approach allows one to study the generation of features in the primordial spectra by parametrizing any nontrivial dynamics containing tiny time-scales into background couplings, such as the sound speed $c_s$ and the expansion rate $H$ appearing in the EFT Lagrangian for the fluctuations. This is precisely the case of both $P(X,\phi)$ models and multi-field models with heavy fields (second regime already mentioned), where the emergence of features in the primordial spectra may be traced back to sudden time variations of the sound speed $c_s$ and the expansion rate $H$ (and subsequent time derivatives) appearing in the effective Lagrangian describing the dynamics of curvature perturbations.

\subsection{Correlating features in the primordial spectra}

The purpose of this article is to study the emergence of features in the primordial spectra by employing the effective field theory of inflation formalism. Adopting this framework will allow us to uncover model independent relations between features appearing in different $n$-point correlation functions of primordial curvature perturbations, in particular between features in the power spectrum and the bispectrum. This will turn out to be possible because the EFT action of the curvature perturbation depends only on a reduced number of background parameters present at every order of its non-linear perturbative expansion. In this way, by relating the time-variation of these parameters, we will be able to deduce a set of consistency relations correlating features in the inflationary primordial spectra.

For instance, it is already well known how to express the bispectrum as a function of the power spectrum in the particular case of features due to small sudden changes of the sound speed $c_s$ of curvature perturbations away from the value $c_s = 1$~\cite{Achucarro:2012fd} (see also ref.~\cite{Achucarro:2014msa}). The existence of such a relation is allowed by the possibility of expressing the sound speed as a function of the power spectrum in the following way
\be
1 - c_s^2(\tau) = \frac{2 }{\pi i}  \int  \frac{d k}{k}  \frac{\Delta \mathcal P}{\mathcal P_0 } (k)  e^{- 2 i k \tau} ,  \label{invert-1}
\ee
where $\tau$ is the usual conformal time, and the difference $\Delta \mathcal P \equiv \P - \P_0$ parametrizes features in the power spectrum $\mathcal P(k)$ against its featureless counterpart $\P_0(k)$ (which has a slight red tilt). This equation results from analyzing the relation determining the power spectrum $\mathcal P(k)$ as a function of $c_s(\tau)$ and inverting it. On the other hand, given that the combination \mbox{$1 - c_s^2$} appears as a coupling in the cubic part of the EFT action for the fluctuations (generating non-Gaussianity), having $c_s$ as a function of the power spectrum permits one to compute the bispectrum in terms of the power spectrum. Specifically, one finds that the $f_{\rm NL}$ parameter\footnote{The $f_{\rm NL}$-parameter is defined in eq.~(\ref{defnl}).} has the general dependence on the power spectrum given by
\be
f_{\rm NL} (k_1 , k_2 , k_3) = \left[ f_2  \frac{d^2}{d \ln k^2}    \frac{\Delta \P}{\P_0} (k)  + f_1  \frac{d}{d \ln k}  \frac{\Delta \P}{\P_0} (k)  + f_0 \frac{\Delta \P}{\P_0} (k) \right]_{k = (k_1 + k_2 + k_3)/2} , \label{general-fNL-features}
\ee
where $f_i = f_i(k_1 , k_2 , k_3)$ are known smooth scale-independent functions of the triangle configuration $\k_1 + \k_2 + \k_3 =0$, determined by the fact that features come from sound speed variations~\cite{Achucarro:2012fd}. Equation~(\ref{general-fNL-features}) is found to reduce to the well known Maldacena's consistency relation~\cite{Maldacena:2002vr, Creminelli:2004yq, Cheung:2007sv} in the squeezed limit, where one of the momenta is much smaller than the other two (\emph{i.e.} $k_3 \ll k_1,k_2$). Moreover, in the particular case of sharp features, where the sound speed is subject to rapid variations with a characteristic time-scale much smaller than $H^{-1}$, the features become dominated by the second order $\ln k$-derivative, giving us back the following general expression for sharp features
\be
 f_{\rm NL}  \simeq  \beta  \left[ \frac{d^2}{d \ln k^2}  \frac{\Delta \mathcal P}{\mathcal P_0} (k)  \right]_{k = (k_1 + k_2 + k_3)/2} , \label{fnl-beta-intro}
\ee
where $\beta = \beta (k_1 , k_2, k_3 )$ is a smooth function, independent of the overall scale, but dependent on the shape of interest and the source of the feature ($f_2$ in the parametrization of eq.~(\ref{general-fNL-features})).

It is possible to generalize eq.~(\ref{invert-1}) to the case in which features are generated by both, time variations in the sound speed and time variations in the expansion rate $H$. This ge\-ne\-ra\-li\-za\-tion was derived in ref.~\cite{Palma:2014hra}, and is found to be given by
\bea
\frac{1}{8} (1 - c_s^2)'''' +  \frac{ \delta_H'' }{ 2Ê\tau^2}  - \frac{ \delta_H}{ \tau^4}  = \frac{4 }{\pi i}  \int_{-\infty}^{+\infty} \!\!\! d k \, k^3 \frac{\Delta \mathcal P}{\mathcal P_0 } (k)  e^{- 2 i k \tau} ,\label{invert-2}
\eea
where $\delta_H \simeq -\frac{1}{2} \tau \eta'$, and $\eta = \dot \epsilon / H \epsilon$ (with $\epsilon = - \dot H / H^2$ the usual slow roll parameter and $'$ standing for derivatives with respect to conformal time $\tau$). However, because both $1 - c_s^2$ and $\eta$ are couplings appearing in the cubic action, it is clear that under more general backgrounds the bispectrum cannot be expressed uniquely in terms of the features appearing in the power spectrum, inevitably introducing a degeneracy in the parameter space determining the source of features.\footnote{This point was specially emphasized in ref.~\cite{Gong:2014spa} where a general expression for the bispectrum in terms of the power spectrum was obtained.} Despite of this, eq.~(\ref{invert-2}) may be used to express the bispectrum as a function of the power spectrum in two limiting cases: when features are generated by a variation of the sound speed, and when they are generated by a variation of the expansion rate. To be more specific, in the case of sharp features, if features are generated by a variation of the sound speed one finds that the bispectrum has the form shown in eq.~(\ref{fnl-beta-intro}) with $\beta$ given by
\be
\beta_\textrm{s} = \frac{5}{6} \frac{k_1 k_2 k_3}{k_1^3 + k_2^3 + k_3^3}    \frac{ k_1^2 + k_2^2 + k_3^2 }{(k_1 + k_2 + k_3)^2} ,  \label{beta-s}
\ee
whereas if features are generated by a variation of the expansion rate, then $\beta$ acquires the form given by:
\be
\beta_{\epsilon} =  \frac{5}{12} \frac{k_1 k_2 k_3}{k_1^3 + k_2^3 + k_3^3} .  \label{beta-epsilon}
\ee
These two results for $\beta$ may in principle be tested by searches of features in the CMB non-Gaussian angular spectrum.

\subsection{Article's main idea \& results}

Equation~(\ref{fnl-beta-intro}), together with the functions $\beta_\textrm{s}$ and $\beta_{\epsilon}$ of eqs.~(\ref{beta-s}) and (\ref{beta-epsilon}), parametrizes features in the two limiting cases in which these are generated by a variation of the sound speed and the expansion rate respectively. From these results, it is reasonable to expect that eq.~(\ref{fnl-beta-intro}) continues to be valid under more general circumstances in which these two sources occur simultaneously, as expected in more realistic models. As we will show in this article, this is indeed the case: there exists a natural one-parameter realization for the function $\beta$ in eq.~(\ref{fnl-beta-intro}) such that $\beta_\textrm{s}$ and $\beta_{\epsilon}$ of eqs.~(\ref{beta-s}) and~(\ref{beta-epsilon}) constitute the desired limiting cases. This one-parameter realization of $\beta$ is a consequence of a relation between $\eta$, parameterizing the evolution of the expansion rate, and the sound speed $c_s$ given by
\be
\eta = \eta_0 + \frac{\alpha}{2}  \tau \frac{d }{d \tau} c_s^2,  \label{newrel}
\ee
where $\alpha$ is a constant specified by the model underlying the emergence of features, and $\eta_0$ represents the slowly varying part of $\eta$. Although it is clear that this non-trivial relation is not guaranteed to be valid in general, we will argue that it remains a fairly good approximation for a large variety of models admitting sudden time variations of their slowly varying background quantities, including $P(X,\phi)$ and multi-field models with heavy fields. For the purposes of the present work, admitting the validity of eq.~(\ref{newrel}) enables us to interpolate between the two limiting cases shown in eqs.~(\ref{beta-s}) and~(\ref{beta-epsilon}), obtaining a much more general relation for $\beta$ given as:
\be
\beta_\alpha =  \frac{5}{12} \frac{1}{1+\alpha} \frac{k_1 k_2 k_3}{k_1^3 +k_2^3 +k_3^3}  \left[ \alpha +2  \frac{ k_1^2+k_2^2+k_3^2 }{(k_1+k_2+k_3)^2} \right]. \label{beta-alpha}
\ee
This relation gives us back eq.~(\ref{beta-s}) in the specific case of $\alpha = 0$ and eq.~(\ref{beta-epsilon}) in the case $|\alpha| \to + \infty$. Equation~(\ref{beta-alpha}) constitutes the main result of this work. It consists of a single parameter consistency relation between the power spectrum and the bispectrum valid for triangle configurations away from the squeezed limit, where one recovers Maldacena's consistency relation. It offers a concrete ---and fairly model independent--- parametrization of features to be tested in future primordial spectra reconstructions using cosmic microwave background and/or large scale structure observations~\cite{Hunt:2013bha, Meerburg:2013cla, Achucarro:2013cva, Meerburg:2014kna, Hazra:2014jwa, Hu:2014hra, Fergusson:2014hya, Munchmeyer:2014cca, Fergusson:2014tza, Gariazzo:2015qea}.

\subsection{Outline}

The present article is organized as follows: We will start in Section~\ref{sec:general-param}, where we propose the use of eq.~(\ref{newrel}) as a valid prescription relating the time dependence of both the sound speed and the inflationary expansion rate. There we show that by admitting this relation one obtains eq.~(\ref{beta-alpha}), linking together the features in the primordial bispectrum and the primordial power spectrum. In Section~\ref{sec:examples} we justify the use of eq.~(\ref{newrel}) and examine its validity within different field theoretical realizations of inflation involving the presence of a sound speed. To be more precise, we numerically solve inflationary backgrounds with sudden features in various classes of $P(X,\phi)$ and multi-field models of inflation. Then, in Section~\ref{sec:resonant-features} we show that our proposed relation~(\ref{newrel}) also serves to parametrize resonant features. Finally, in Section~\ref{sec:conclusions} we provide our concluding remarks on the findings of this work.

\section{A general parametrization of features}  \label{sec:general-param}
\setcounter{equation}{0}

The appearance of sharp features in the primordial spectra may be traced back to the time-variation of background quantities at a rate much larger than those characteristic of slow-roll inflation~\cite{Chen:2006xjb, Chen:2008wn, Arroja:2011yu, Adshead:2011bw, Adshead:2011jq, Arroja:2012ae}. To analyze the effect of these rapid variations on the dynamics of fluctuations, we adopt the EFT of inflation perspective. This framework asserts that the action describing the dynamics of curvature perturbations may be parametrized with the help of a limited number of background parameters, determining non-trivial relations between coefficients appearing at different orders in the theory. In comoving gauge, the action for primordial curvature perturbations, up to cubic order, may be written in the following way
\be
S = S^{(2)} + S^{(3)}, \label{basic-S-def}
\ee
where the quadratic part of the action, $S^{(2)}$, corresponds to (written in units such that $m_{\rm Pl}^2=1$)
\be
S^{(2)} =  \int \! d^4 x \,  a^3 \epsilon \left[ \frac{1}{c_s^2} \dot {\mathcal{R}}^2 - \frac{1}{a^2}(\nabla \mathcal{R})^2  \right] , \label{R-2-action}
\ee
whereas the cubic part, $S^{(3)}$, is given by~\cite{Burrage:2011hd}
\bea
S^{(3)} &=&  \int \! d^4 x \,  a^3 \epsilon \bigg[ 
\frac{1}{c_s^4} \left[ 3 (c_s^2 - 1) + \epsilon - \eta \right] \R \dot \R^2 + \frac{1}{c_s^2 a^2} \left( (1 - c_s^2) + \eta + \epsilon -  \frac{2 \dot c_s}{H c_s}  \right) \R (\nabla \R)^2  \nn \\
&&  + \frac{1}{H }  \left(\frac{1 - c_s^2}{c_s^4}  - \frac{ 2 \lambda }{ \epsilon H^2} \right)  \dot \R^3  + \frac{1}{4 a^4} (\partial \chi)^2 \nabla^2 \R - \frac{4 - \epsilon}{2 \epsilon a^4} \nabla^2 \chi \partial^i  \R \partial_i \chi + \frac{f }{\epsilon a^3} \frac{\delta S^{(2)}}{\delta \R}
 \bigg] , \label{S-R-3-second}
\eea
where $\chi$ is given by the constraint equation $\nabla^{2} \chi = a^2 \epsilon \dot \R /  c_s^{2}$. In these expressions, the expansion rate $H$ and the slow roll parameters $\epsilon$ and $\eta$ parametrize the evolution of the scale factor $a(t)$ as
\be
H = \frac{\dot a}{a} , \qquad \epsilon = - \frac{\dot H}{H^2} , \qquad  \eta =  \frac{\dot \epsilon}{H \epsilon} , \label{def-slow-roll}
\ee
where the dot represents a derivative with respect to cosmic time $t$. A relevant background quantity parametrizing coefficients appearing in both quadratic and cubic parts of the action, corresponds to the sound speed $c_s(t)$ which determines the speed at which long wavelength modes propagate. On the other hand, the parameter $\lambda$ in eq.~(\ref{basic-S-def}) parametrizes the strength of the operator $\dot \R^3$, and it is usually found to depend on $c_s$ according to a relation determined by the specific model in question. Finally, the quantity $f$ multiplying the linear classical equation of motion $\delta S^{(2)} / \delta \R$ is a given quadratic function of $\R$, whose specific form will turn out to be irrelevant for the present discussion.

This action may be simplified substantially by making a few reasonable assumptions about the type of features we want to study. First of all, we will consider variations of the sound speed $c_s$ such that it stays close to the value $c_s = 1$. That is, we will assume that 
\be
\theta   \equiv 1 - c_s^2 \ll 1 ,
\ee
at all times. We find this assumption reasonable, because a suppressed sound speed $c_s$ is known to produce a sizable level of equilateral non-Gaussianity~\cite{Seery:2005wm, Chen:2006nt}, so far undetected~\cite{Ade:2013ydc}. An immediate consequence of this choice relates to the size of the coefficient in front of the operator $\dot \R^3$ appearing in eq.~(\ref{S-R-3-second}). Indeed, from an effective field theory point of view, it is possible to show that this coefficient is in fact \emph{naturally} of order $\theta^2$ (see for instance, ref.~\cite{Senatore:2009gt}):
\be
\frac{1-c_s^2}{c_s^4} - \frac{2 \lambda}{\epsilon H^2} \sim  \frac{(1 - c_s^2)^2}{c_s^4}  \sim \theta^2.
\ee
Another important assumption that we will adopt is that the time variation of the background does not break the quasi de Sitter phase characterizing inflation during the horizon crossing of the modes responsible for the primordial spectra accessible to observations. In other words, we will assume that the evolution of $H$ is such that
\be 
\epsilon \ll 1, \label{quasi-de-sitter-cond}
\ee
at all times. In practice, because $\epsilon$ is restricted to be positive, eq.~(\ref{quasi-de-sitter-cond}) implies that $\epsilon$ will have the form 
\be
\epsilon = \epsilon_0 + \Delta \epsilon ,
\ee
where $\epsilon_0$ corresponds to the slowly varying part of $\epsilon$, characterizing the average slow-roll behavior of inflation, and $\Delta \epsilon$ is the part containing rapid departures from $\epsilon_0$. In addition to condition (\ref{quasi-de-sitter-cond}), we will assume that $\Delta \epsilon$ is such that its variation is small relative to $\epsilon_0$. In other words:
\be
\Delta \epsilon \ll \epsilon_0.
\ee
This condition means that the background of our interests contains rapid variations, but will continue to stay close to the quasi-de Sitter stage parametrized by the featureless value $\epsilon_0$. The previous splitting of $\epsilon$ implies that $\eta$ may be written in terms of $\Delta \epsilon$ in the following way 
\be
\eta =  \eta_0  + \Delta \eta , \qquad  \Delta \eta \equiv \frac{\Delta \dot \epsilon}{H \epsilon_0 }  ,  \label{def-slow-roll-2}
\ee
where $\eta_0 = \dot \epsilon_0 / H \epsilon_0$ represents the slowly varying part of $\eta$. Finally, we will assume that the time variation of both $\Delta \epsilon$ and $c_s$ are characterized by a time scale much smaller than $H^{-1}$ (which is what we mean by rapid variations). Consequently, the features we are interested in are characterized by the following hierarchies:
\be
\frac{1}{H} \Big| \frac{d \theta}{dt} \Big| \gg  \theta ,  \qquad
\frac{1}{H} \Big| \frac{d \Delta \epsilon}{dt} \Big| \gg  \Delta \epsilon . \label{var-Delta-epsilon}
\ee
It may be appreciated that condition~(\ref{var-Delta-epsilon}) implies that $\Delta \eta$, defined in~(\ref{def-slow-roll-2}), may be of order $1$ (or even larger) without implying a breaking of condition (\ref{quasi-de-sitter-cond}). More generally, we assume that any background quantity $A$ parametrizing rapid variations (such as $\theta$, $\Delta \epsilon$ and $\Delta \eta$) satisfies:
\be
\frac{1}{H} \Big| \frac{d A}{dt} \Big| \gg | A | .  \label{hierarchy-B-1}
\ee 
It is also useful to cast this hierarchy in terms of both conformal time $\tau$ and $e$-folds $N$. Let us recall that conformal time is defined to satisfy the relation $d \tau = d t / a$, whereas $N$ satisfies $d N = H dt$. Then, the previous hierarchy may be written in the following two alternative ways, that will turn out to be useful in the next discussions:
\be
 \Big| \tau \frac{d A}{d\tau} \Big| \gg | A | , \qquad  \Big| \frac{d A}{dN} \Big| \gg | A | . \label{hierarchy-B-2}
\ee 
To continue, all of the previous assumptions may be put together to organize and simplify the form of our initial action given in eq.~(\ref{basic-S-def}). In particular, the cubic part of the action is found to be dominated by the following terms:
\bea
S^{(3)}_{\rm int} = - \int \! d^4 x \,  a^3 \epsilon_0  \bigg\{   (3 \theta + \eta) \R \dot \R^2 + \frac{1}{a^2} (  \tau \theta' - \eta)  \R (\nabla \R)^2   \bigg\} .  \label{final-S3-int}
\eea
This result allows us to study the effects of features in the bispectrum using the in-in formalism of cosmological perturbation theory. Details of how this is done may be found in ref.~\cite{Palma:2014hra}. Here instead we quote the main results of Appendix~\ref{app1} in order to derive various relevant expressions to obtain the primordial spectra.

\subsection{Features in the power spectrum}

Let us first consider the effects of a rapid variation of the background on the power spectrum. The dimensionless power spectrum $\P(k)$ is usually defined in terms of the 2-point correlation function of curvature perturbations in Fourier space (evaluated at the end of inflation) as:
\be
\langle \R_{\k} \R_{\k'} \rangle \equiv (2 \pi)^3 \delta(\k + \k') \frac{2 \pi^2}{k^3} \P(k) .
\ee
The computation of $\langle \R_{\k} \R_{\k'} \rangle$ may be performed using the in-in formalism of perturbation theory discussed in Appendix~\ref{app1}. In particular, one finds that the power spectrum $\P(k)$ has the following form (see Appendix \ref{app1}):
\be
\P (k)=\P_0 + \Delta \P (k), \qquad \P_0 =\frac{H_0^2}{8\pi^2\epsilon_0} \left( \frac{k}{k_*} \right)^{n_s - 1}. \label{Pofk}
\ee
Here $\P_0$ corresponds to the zeroth order featureless power spectrum, which coincides with the conventional prediction offered by canonical single field slow-roll inflation ($n_s$ is the spectral index parametrizing the small scale dependence of the power spectrum, and $k_*$ is a pivot scale fixed by mode that exited the horizon when the background was characterized by an expansion rate $H_0$). On the other hand, $\Delta \P$ is the part containing the features, and is given by~\cite{Palma:2014hra}
\bea
\frac{\Delta \mathcal P}{\mathcal P_0 } (k) &=&  k \int_{-\infty}^{0} \!\!\! d \tau \, \left[  - \theta + \frac{\delta_H}{k^2Ê\tau^2}  + \frac{2 \delta_H}{k^4Ê\tau^4}  - \frac{1}{k^4Ê\tau^3} \frac{d \delta_H}{d \tau} \right] \, \sin (2 k \tau) , \label{features-delta}
\eea
where $\tau$ represents conformal time. In addition, $\delta_H$ is a function containing information about the time varying background, which in the case of rapid variations is found to be given by~\cite{Palma:2014hra}:
\be
\delta_H = -  \frac{1}{2} \tau \eta' . \label{delta-H-eta}
\ee
We may quickly verify the validity of eq.~(\ref{features-delta}) for the particular case in which $\theta =\theta_0 \ll 1$ is a constant and $\delta_H=0$. In this case the integral of eq.~(\ref{features-delta}) may be solved by dropping the highly oscillatory part of the lower limit $\tau \to - \infty$, giving us back
\bea
\frac{\Delta \mathcal P}{\mathcal P_0 } =   \frac{\theta_0}{2} =  \frac{1 - c_s^2}{2} \simeq 1 - c_s. \label{check-constant-speed}
\eea
This result, together with eq.~(\ref{Pofk}) gives us back the result $\P = \P_0 / c_s = H_0^2 / 8\pi^2\epsilon_0 c_s$, which is the well known expression valid for the power spectrum for curvature perturbations with a constant sound speed.

To continue, eq.~(\ref{features-delta}) may be simplified by assuming that both $\theta \to 0$ and $\delta_H \to 0$ as $\tau \to 0^{-}$, which simply means that small variations of the background happening at the end of inflation are meaningless for the purpose of computing features in the long wavelength power spectrum, describing modes that exited the horizon long before inflation finishes. This assumption allows one to perform several integrations by part and write:
\be
k^3 \frac{\Delta \mathcal P}{\mathcal P_0 } (k) = - \frac{1}{2} \int_{-\infty}^{0} \!\!\! d \tau \, \left[ \frac{1}{8} \theta'''' + \frac{ \delta_H'' }{ 2Ê\tau^2}  - \frac{ \delta_H}{ \tau^4}   \right] \, \sin (2 k \tau) \, . \label{features-delta-2}
\ee
Let us emphasize here that up to this point, we have made no assumptions about the sharpness of features. This new expression may be Fourier-inverted by conveniently extending the time domain of the functions $\theta$ and $\delta_H$ to the whole range $\tau \in (- \infty , + \infty)$, and demanding them to be odd under $\tau \to - \tau$. The result is found to be given by eq.~(\ref{invert-2}) already shown in the introduction. However, if we further use the fact that $\delta_H$ satisfies the hierarchy of eq.~(\ref{hierarchy-B-2}), we are allowed to drop the lower order time-derivatives from the left hand side of eq.~(\ref{invert-2}), from where we finally obtain:
\be
\frac{1}{8} \theta'''' -  \frac{ \eta''' }{ 4Ê\tau}    = \frac{4 }{\pi i}  \int_{-\infty}^{+\infty} \!\!\! d k \, k^3 \frac{\Delta \mathcal P}{\mathcal P_0 } (k)  e^{- 2 i k \tau} . \label{inverted-power-2}
\ee
This relation gives us the time dependence of a specific combination of background quantities in terms of the features appearing in the power spectrum. Before moving on to consider features in the bispectrum, it is important to notice that the hierarchy of eq.~(\ref{hierarchy-B-2}) together with eq.~(\ref{inverted-power-2}) implies a hierarchy between $k$-derivatives for the power spectrum, which  takes the form:
\be
\left| \frac{d^2}{d \ln k^2} \frac{\Delta \P}{\P_0}  \right| \gg \left| \frac{d}{d \ln k} \frac{\Delta \P}{\P_0}  \right| \gg \left| \frac{\Delta \P}{\P_0}  \right| . \label{features-hierarchy}
\ee
In other words, rapid variations of the background (characterized by a timescale much smaller than $H^{-1}$) imply sharp features in the power spectrum. We will use this important result in the next discussions, and verify its validity in Section~\ref{sec:examples}, where we examine several concrete examples of models with features.

\subsection{Features in the bispectrum}

We may now examine the appearance of features in the bispectrum. Let us recall that the bispectrum $B$ may be defined through the 3-point correlation function of curvature perturbations, in Fourier space, at the end of inflation as
\be
 \langle \hat \R_{\k_1}  \hat \R_{\k_2}  \hat \R_{\k_3}  \rangle = (2 \pi)^3 \delta (\k_1 + \k_2 + \k_3 ) B (\k_1 , \k_2 , \k_3 ) ,
\ee
where, because of the delta function $\delta (\k_1 + \k_2 + \k_3 )$, the three momenta are restricted to add up to zero. By computing the three point correlation function using the in-in formalism one finally obtains
\be
B =B_0+\Delta B,
\ee
where $B_0$ represents the standard featureless, slow-roll suppressed bispectrum~\cite{Chen:2006nt}. On the other hand, $\Delta B$ is the part containing features, and is found to be given by (see Appendix \ref{app1})
\bea
\Delta B (\k_1, \k_2, \k_3)   \! &=& \!   
  \frac{2 \pi^4 \P_0^2}{(k_1 k_2 k_3)^{3}}\int_{-\infty}^\infty d \tau ~i ~e^{i K \tau} \times\nn\\
&& ~ \Bigl[  (3\theta+\eta)    \left[ -(k_2k_3)^2-(k_3k_1)^2-(k_1k_2)^2 + i k_1 k_2 k_3 \tau (k_2k_3+k_3k_1+k_1k_2)   \right]         \nn\\
&& ~  +  \frac{\eta-\tau\theta' }{2\tau^2} \left[k_1^2+k_2^2+k_3^2  \right] (1-ik_1\tau) (1-ik_2\tau) (1-ik_3\tau)             \Bigr], \label{Delta-B}
\eea
with the shorthand $K = k_1 + k_2 + k_3$. This relation is equivalent to that of eq.~(\ref{features-delta}) and allows us to compute the shape and size of features in the bispectrum out from the varying background quantities $\theta$ and $\eta$. 

\subsection{A proposal to relate background quantities} \label{sec:proposal}

Crucially, both $\Delta \P$ and $\Delta B$ depend on $\theta$ and $\eta$ simultaneously. This implies that $\Delta \P$ and $\Delta B$ are in principle not tied together, and may contain scale dependent features which do not correlate with one another. Nevertheless, we may break the degeneracy in the pa\-ra\-me\-ter space of these observables by assuming a dynamical relation between the background quantities $\theta$ and $\eta$. In this respect, our proposal consists of assuming the following relation
\be
\eta = \eta_0 -\frac{\alpha}{2} \tau \theta', \label{new_rel}
\ee
where $\eta_0 = \dot \epsilon_0 / H \epsilon_0$, and $\alpha$ is a slowly varying dimensionless function that may be considered to be a constant for all practical purposes.  We will justify the use of this relation in the next section, where we will verify its validity within a wide range of models, including $P(X,\phi)$ and multi-field models of inflation. In the meantime, notice that this relation, together with eq.~(\ref{features-delta-2}), implies that the power spectrum is now given by:
\bea
\frac{\Delta \mathcal P}{\mathcal P_0 } (k) &=& - (1+\alpha) k \int_{-\infty}^{0} \!\!\! d \tau \,  \theta (\tau) \, \sin (2 k \tau) . \label{psapp}
\eea
On the other hand, using eq.~(\ref{inverted-power-2}) allows us to determine the form of both $\theta = 1 - c_s^2$ and $\Delta \eta = \eta - \eta_0$ in terms of the features appearing in the power spectrum as:
\bea
\theta &=& \frac{1}{1+\alpha}\frac{2 }{\pi i}  \int_{-\infty}^{+\infty} ~\!\!\! \frac{d k}{k}~ \frac{\Delta \mathcal P}{\mathcal P_0 } (k)  e^{- 2 i k \tau} , \label{theta-Power} \\ 
\Delta \eta &=&\frac{\alpha}{1+\alpha} \frac{1}{\pi i}  \int_{-\infty}^{+\infty} \!\!\! d k \left( \frac{\partial}{\partial k}   \frac{\Delta \mathcal P}{\mathcal P_0 } (k) \right) e^{- 2 i k \tau} .  \label{eta-Power}
\eea
It may be seen that in the limit $\alpha \to 0$ one recovers the case in which the features are exclusively due to variations of the sound speed, and in the limit $|\alpha| \to + \infty$ one reobtains the case in which features are due to variations of the expansion rate. One may worry about the limit $\alpha \to -1$, which makes eqs.~(\ref{theta-Power}) and~(\ref{eta-Power}) diverge. In that limit, subleading terms that were neglected at the left hand side of eq.~(\ref{inverted-power-2}) take over, and modify the final forms of eqs.~(\ref{theta-Power}) and~(\ref{eta-Power}). In the present analysis we omit this specific situation and leave the related problem of finding such expressions open (we further comment on this issue in Section~\ref{accuracy-power}).

\subsection{A general correlation between features}

Now that we have both $\theta$ and $\eta$ expressed in terms of the piece $\Delta \P$ determined by features, we may use eq.~(\ref{Delta-B}) to reach a general expression determining $\Delta B$ as a function of $\Delta \P$. To obtain it, it is enough to replace eqs.~(\ref{theta-Power}) and~(\ref{eta-Power}) back into eq.~(\ref{Delta-B}). To write the result, it is convenient to introduce the standard $f_{\rm NL}$-parameter as a function of $k_1$, $k_2$, and $k_3$, defined as:
\be
 f_{\rm NL} \equiv \frac{10}{3} \frac{k_1 k_2 k_3 }{k_1^3 + k_2^3 + k_3^3} \frac{(k_1 k_2 k_3)^2 }{(2 \pi)^4 \P_0^2} \Delta B .  \label{defnl}
\ee
Then, by keeping the leading contributions to $\Delta B$, which according to eq.~(\ref{features-hierarchy}) consists of a term proportional to second order derivatives of $\Delta \P$ in terms of $\ln k$ (and a term proportional to a first order derivative that dominates in the squeezed limit) we finally find
\be
 f_{\rm NL}  \simeq  \left[  \beta_\alpha  \frac{d^2}{d \ln k^2}  \frac{\Delta \mathcal P}{\mathcal P_0 } (k)  -\frac{5}{12}  \frac{d}{d \ln k} \frac{\Delta \P}{\P_0}(k) \right] _{k = (k_1 + k_2 + k_3)/2} , \label{f_NL_features}
\ee
where $\beta_\alpha = \beta_{\alpha} (k_1 , k_2 , k_3)$ is a scale-independent function given by:
\be
\beta_\alpha = \frac{5}{12}\frac{1}{1+\alpha}  \frac{k_1 k_2 k_3}{k_1^3 +k_2^3 +k_3^3}  \left[ \alpha + 2 \frac{ k_1^2+k_2^2+k_3^2 }{(k_1+k_2+k_3)^2} \right] .  \label{resu}
\ee
This final form of $\beta$ interpolates between the results (\ref{beta-s}) and (\ref{beta-epsilon}) discussed in the introduction, found in the two separate cases in ref.~\cite{Palma:2014hra}. In particular, it is worth highlighting that in the equilateral configuration $k_1\simeq k_2 \simeq k_3$, $\beta$ has the form
\be
\beta_\alpha^{\rm (eq)} = \frac{5}{36}\frac{1}{1+\alpha} \left[\alpha+\frac{2}{3}\right],
\ee
while in the folded configuration $k_1=k_2 = k_3/2$, $\beta$ becomes: 
\be
\beta_\alpha^{\rm (fold)} = \frac{1}{12}\frac{1}{1+\alpha} \left[ \alpha + \frac{3}{4}\right].
\ee
Finally, it is worth noticing that in the squeezed limit $(k_1=k_2, k_3\to 0)$ we recover the well known consistency relation (see also ref.~\cite{Sreenath:2014nca}), as the coefficient of the second derivative of the power spectrum disappears
\be
f^{\rm (sq)}_{\rm NL} = -\frac{5}{12} \left[ \frac{d}{d \ln k} \frac{\Delta \P}{\P_0}(k) \right] ,
\ee
independently of the value of $\alpha$. As already mentioned in the introduction, eq.~(\ref{resu}) constitutes our most important result. It provides a unique relation determining the shape and size of features in the bispectrum in terms of those appearing in the power spectrum with only one parameter to adjust. As we shall see in the next section, the parameter $\alpha$ is determined by the model responsible for features, and in principle may be constrained by observations aiming to characterize the scale dependence of the the primordial spectra. At any rate, it is important to recall that eq.~(\ref{resu}) is strictly valid away from the value $\alpha=-1$, and that it assumes that features were generated by rapidly varying background quantities satisfying eq.~(\ref{hierarchy-B-2}).

\subsection{On the general structure of the correlated spectra}

The results of this section ratify the general dependence anticipated by eq.~(\ref{general-fNL-features}) of the introduction. A striking aspect of this relation is the fact that $f_{\rm NL}$ does not depend on third order derivatives of $\Delta \P/\P_0$ or even higher. In fact, (\ref{general-fNL-features}) constitutes a truncated version of a more general relation between $f_{\rm NL}$ and derivatives of $\Delta \P/\P_0$ that may be expressed as:
\be
f_{\rm NL} (k_1 , k_2 , k_3) =  \sum_{n=0}^{n = \infty} f_n  \frac{d^n}{d \ln k^n}  \frac{\Delta \P}{\P_0} (k) . \label{most-general-f-P}
\ee
However, it is possible to verify that $f_n$ with $n>2$ are slow-roll suppressed. To appreciate this, first notice that $\ln k$-derivatives acting on $\Delta \P/\P_0$ are the consequence of terms con\-tai\-ning powers of conformal time $\tau$ appearing inside the integral of (\ref{Delta-B}). These in turn, come from the time dependence of the curvature perturbation wave function $\R_k (\tau)$ given in (\ref{R-I}), valid in the de Sitter limit $\epsilon \to 0$ (see the discussion of the Appendix~\ref{app1}). Nevertheless, this specific form of $\R_k (\tau)$ receives corrections of order $\epsilon$ that introduce additional powers of $\tau$ inside (\ref{Delta-B}), finally leading to (\ref{most-general-f-P}) but with coefficients $f_n$ (with $n>2$) suppressed by slow-roll parameters. This observation gives us an upper limit on the sharpness of features that our method is allowed to study. That is, in order for the truncation to remain valid, features in the power spectrum must be such that:
\be
\epsilon \left| \frac{d^3}{d^3 \ln k} \frac{\Delta \P}{\P_0} \right|  \ll \left| \frac{d^2}{d^2 \ln k} \frac{\Delta \P}{\P_0} \right| .
\ee
In other words, our approach is useful to analyze sharp features, but their sharpness is limited by the size of the slow-roll parameter $\epsilon$.

\section{Sound speed and expansion rate during inflation} \label{sec:examples}
\setcounter{equation}{0}

Our main result~(\ref{resu}), deduced in the previous section, depends crucially on the validity of eq.~(\ref{new_rel}) relating the sound speed $c_s$ and the slow-roll parameter $\eta$. In this section we provide arguments to support this conjecture and discuss a few examples for which it is found to be valid. Let us start by emphasizing that, a priori, there are no reasons to expect a simple relation determining the value of $\eta$ in terms of the sound speed $c_s$ since, in principle, the background dynamics has enough degrees of freedom to allow for situations in which $\eta$ and $c_s$ evolve independently. Nevertheless, in the class of models that we are attempting to describe, features appear as the consequence of small deviations from the quasi-de Sitter background driven by inflation, characterized by constant values of $\epsilon$ and $c_s^2$:
\be
\epsilon = \epsilon_0 , \qquad c_s^2 = c_0^2 .
\ee
In these models, $\epsilon$ parametrizes the quasi de Sitter geometry whereas $c_s$ parametrizes the non-trivial kinematical properties of the fluid that is causing the quasi-de Sitter dynamics in the first place. It is therefore reasonable to envisage that a small change in $\epsilon$ away from $\epsilon_0$ will come together with a small compensating change to the sound speed $c_s^2$ away from the value $c_0^2$. In other words, whatever may be the cause of the rapid background variations, it should induce simultaneous variations of $\epsilon$ and $c_s^2$. As long as these variations are small, we expect that
\be
\Delta \epsilon \propto   \Delta c_s^2 , \qquad \Delta \epsilon = \epsilon - \epsilon_0 \ll \epsilon_0, \qquad \Delta c_s^2 = c_s^2 - c_0^2 \ll c_0^2, \label{conjecture}
\ee
where $\epsilon_0$ and $c_0$ are the slowly varying values of $\epsilon$ and $c_s$, that may be taken as constant values for all practical purposes. The idea behind this relation is simple: each time a constant value of $\epsilon$ is reached (that is, each time $\frac{d}{dN}\Delta \epsilon = 0$) a new quasi-de Sitter geometry is achieved, which should be characterized by a new constant value of the sound speed $c_s$. This precisely requires that both quantities vary in synchrony. The proportionality relation $\Delta \epsilon \propto   \Delta c_s^2$ allows us to deduce a more useful relation between $\eta$ and $c_s$. Indeed, by using eq.~(\ref{conjecture}) back in eq.~(\ref{def-slow-roll-2}) where we defined $\Delta \eta$, we obtain the non-trivial relation
\be
\eta = \eta_0 - \frac{\alpha}{2} \frac{d \Delta c_s^2}{dN} , \label{relation-eta-sound}
\ee
where $\alpha$ represents some slowly varying background quantity determined by the specific model causing the rapid variation. Then, in the particular case in which the sound speed varies away from the background value $c_0 =1$, we finally obtain
\be
\eta = \eta_0 + \frac{\alpha}{2} \frac{d \theta}{dN} ,  \label{relation-eta-sound-2}
\ee
which is our desired result. In what follows we check the validity of these results by analyzing several concrete examples.

\subsection{Features in $P(X, \phi)$-models}

For our first two examples we consider a class of $k$-essence models~\cite{ArmendarizPicon:1999rj} with a Lagrangian of the form
$\mathcal{L}= P(X, \phi)$, where $P(X, \phi)$ is a given function of $X= - (\partial \phi)^2/2$ and $\phi$. The homogeneous background equations describing the evolution of inflation are given by the Friedman equation and the equation of motion for the scalar field $\phi$, which, in terms of  number of $e$-folds $N$, are found to be
\bea
3 H^2 - E &=& 0, \label{eq-1-PofX} \\
\frac{d^2 \phi}{dN^2} +\left( 3 c_s^2 - \epsilon \right)\frac{d\phi}{dN} +  \frac{c_s^2}{H^2 P_X } E_\phi &=& 0 ,  \label{eq-2-PofX}
\eea
where now $X = \dot \phi^2 / 2$ (due to the homogeneity of the background), $E \equiv 2 X P_{X} -P$ is the energy density of the inflaton field, and $E_\phi \equiv \partial_\phi E$, $P_X \equiv \partial_X P$, etc. In these models, the slow roll parameter $\epsilon$ and the sound speed $c_s$ are respectively given by:
\be
\epsilon = \frac{X P_X}{H^2} ,  \qquad  c_s^2 = \frac{P_X}{P_X+2XP_{XX}} . \label{quasi-de-sitter-PofX}
\ee
Both of these quantities stay almost constant in a quasi-de Sitter regime. Thus, the origin of features may only appear as the consequence of a non-trivial dependence of these quantities on the rolling inflaton field $\phi$, producing displacements of $\epsilon$ and $c_s^2$ away from their quasi-de Sitter fiducial values $\epsilon_0$ and $c_0^2$. These displacements will happen not only due to the explicit appearance of $\phi$ in eq.~(\ref{quasi-de-sitter-PofX}), but also on the time variations of $X$ responding to this non-trivial dependence. However, as we shall see, as long as these displacements are small, they will happen in synchrony, in such a way that quasi-de Sitter is recovered in terms of both parameters at the same time. In what follows we examine a few examples for which this happens and corroborate the validity of eqs.~(\ref{conjecture})-(\ref{relation-eta-sound-2}).

\subsubsection{Example 1: Localized features in the potential} \label{sec:example-1}

Let us start by considering the specific case in which a rapid variation of the background is caused by a localized feature in the potential. To be concrete, we choose a $P(X, \phi)$-theory of the form
\be
P(X, \phi) = X + A X^2 - V(\phi),
\ee
where $A$ parametrizes a non-trivial contribution to the kinetic term, quadratic in $X$, and $V(\phi)$ corresponds to a chaotic potential with a small feature on top of it in the following manner
\be
V(\phi)=\frac{ m^2
\phi^2}{2} \left[ 1+ f(\phi) \right] ,
\ee
where $m$ is a mass scale parametrizing the featureless chaotic potential, and $f(\phi)$ is a function parametrizing the localized feature. To discuss this class of models let us consider two particular choices for $f(\phi)$. For our first choice, model-$(a)$, we consider a small step feature of the form
\be
f_{(a)}(\phi) = B \tanh\left[\frac{\phi-\phi_0}{\sqrt{2} \Delta \phi}\right] ,
\ee
where $\phi_0$ gives us the position of the step, $\Delta \phi$ its width, and $B$ the size of the jump produced by the step. For our second choice, model-$(b)$, we consider a small Gaussian bump in the potential of the form
\be
f_{(b)}(\phi) = B \exp \left[- \frac{(\phi-\phi_0)^2}{ 2 \Delta \phi^2 }\right] , 
\ee
where, again, $\phi_0$ represents the position of the step, $\Delta \phi$ its width, and $B$ the amplitude of the bump. These potentials have been studied in the work \cite{Cadavid:2015hya}. To be consistent with the hierarchy of eq.~(\ref{hierarchy-B-2}) we assume that both features are ``sharp" in the sense that their effects on the solutions $\phi(N)$ and $H(N)$ take place within an $e$-fold of inflation. In practice, this means that $\Delta \phi $ should be such that $\Delta \phi \ll |d \phi / d N|$. In this way, even if we choose the amplitude $B$ of the features such that they have a tiny effect on $\epsilon$ and $c_s^2$, the sharpness due to the small value of $\Delta \phi $ may induce a dramatic effect on $\eta$ and $\partial_N \theta$. More to the point, let us consider the following values (in Planck units) for the parameters in both models:\footnote{In all our numerical examples, we define the end of inflation by $\eps=0.1$, and choose our parameters such that $\P_0=2.43\cdot 10^{-9}$ $60$ e-folds before the end of inflation. The feature takes place inside the $CMB$-window: in all plots the point $N=0$ corresponds to $55$ e-folds before the end of inflation.}
\be
A=10^{10}, \quad \phi_0=14.19, \quad  B= 5 \times 10^{-4}, \quad \Delta \phi=0.002, \quad m=5.975 \times 10^{-6}. \label{values-parameters-example1}
\ee
These parameter-values imply a background sound speed $c_0^2 \simeq 0.8$. It is important to assert that the formulas of the previous sections ---involving the primordial spectra--- are strictly valid for the case $c_0^2 = 1$, and therefore we are not allowed to use this example to infer a relation between the bispectrum and the power spectrum as discussed in Section~\ref{sec:general-param}. In spite of this fact, this example is still useful to reinforce our confidence on the validity of eqs.~(\ref{conjecture}) and~(\ref{relation-eta-sound}). We have chosen the value of $\phi_0$ in such a way that it produces a feature about 60 $e$-folds before the end of inflation, roughly implying the generation of features in the power spectrum within the window of scales relevant for CMB observations.

\begin{figure}[t!]
\begin{center}
\includegraphics[
width=1.0\textwidth,
  ]{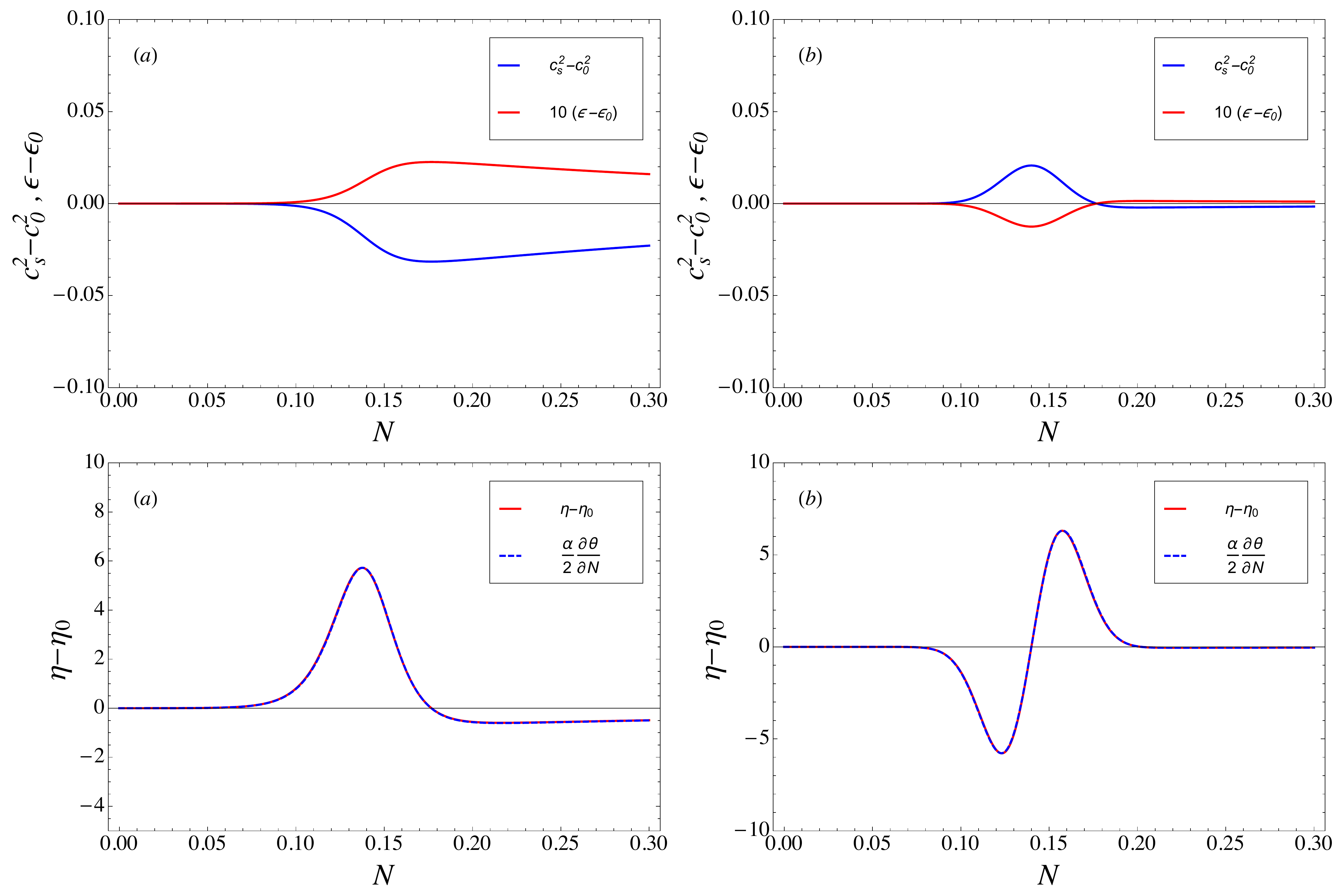} 
\end{center}
\caption{\footnotesize  The figure shows the main numerical results for model-$(a)$ (left-panels) and model-$(b)$ (right-panels). The top panels show the functions $\Delta \epsilon = \epsilon - \epsilon_0$ and $\Delta c_s^2 = c_s^2 - c_0^2$ as a function of $N$. The bottom panels show $\Delta \eta$ and $\alpha \partial_N \theta / 2$ as a function of $N$. Notice that to satisfy eq.~(\ref{relation-eta-sound}), model-$(a)$ requires $\alpha \simeq 15$ whereas model-$(b)$ requires $\alpha \simeq 15.6$.} \label{fig:prel}
\end{figure}

Figure~\ref{fig:prel} summarizes the main numerical results obtained by solving the equations of motion (\ref{eq-1-PofX}) and (\ref{eq-2-PofX}) for the two models in question. The left hand side panels show the results pertinent to model-$(a)$, whereas the right hand side panels show the results for model-$(b)$. In particular, the top panels show the functions $\Delta \epsilon = \epsilon - \epsilon_0$ and $\Delta c_s^2 = c_s^2 - c_0^2$ as a function of $N$. On the other hand, the bottom panels compare $\Delta \eta$ and $\alpha \partial_N \theta / 2$ as a function of $N$ for the values 
\be
\alpha_{(a)} \simeq 15, \qquad \alpha_{(b)} \simeq 15.6 ,
\ee
for models $(a)$ and $(b)$ respectively.
We see that the plots ratify, in a rather eloquent way, the validity of eqs.~(\ref{conjecture}) and~(\ref{relation-eta-sound}). The features on the potential with the parameters values of eq.~(\ref{values-parameters-example1}) imply the rapid variation of background quantities within a window of $\Delta N \sim 0.075$ $e$-folds. In both examples the variations experienced by $\Delta \epsilon = \epsilon - \epsilon_0$ and $\Delta c_s^2 = c_s^2 - c_0^2$ remain small, but the variations of $\Delta \eta$ and $\partial_N \theta$ are found to be large. Moreover, the fact that $\alpha_{(a)}$ and $\alpha_{(b)}$ are larger than $1$ implies that the features are mostly the result of the rapid variation of the expansion rate. We have tried values for the parameters different from those of eq.~(\ref{values-parameters-example1}) and have found similar results, except for the cases in which conditions $\Delta \epsilon \ll \epsilon_0$ and $|\Delta c_s^2 | \ll c_0^2 $ are violated.

\subsubsection{Example 2: DBI inflation with features} \label{example-DBI}

\begin{figure}[t!]
\begin{center}
\includegraphics[width=1.0\textwidth,
]{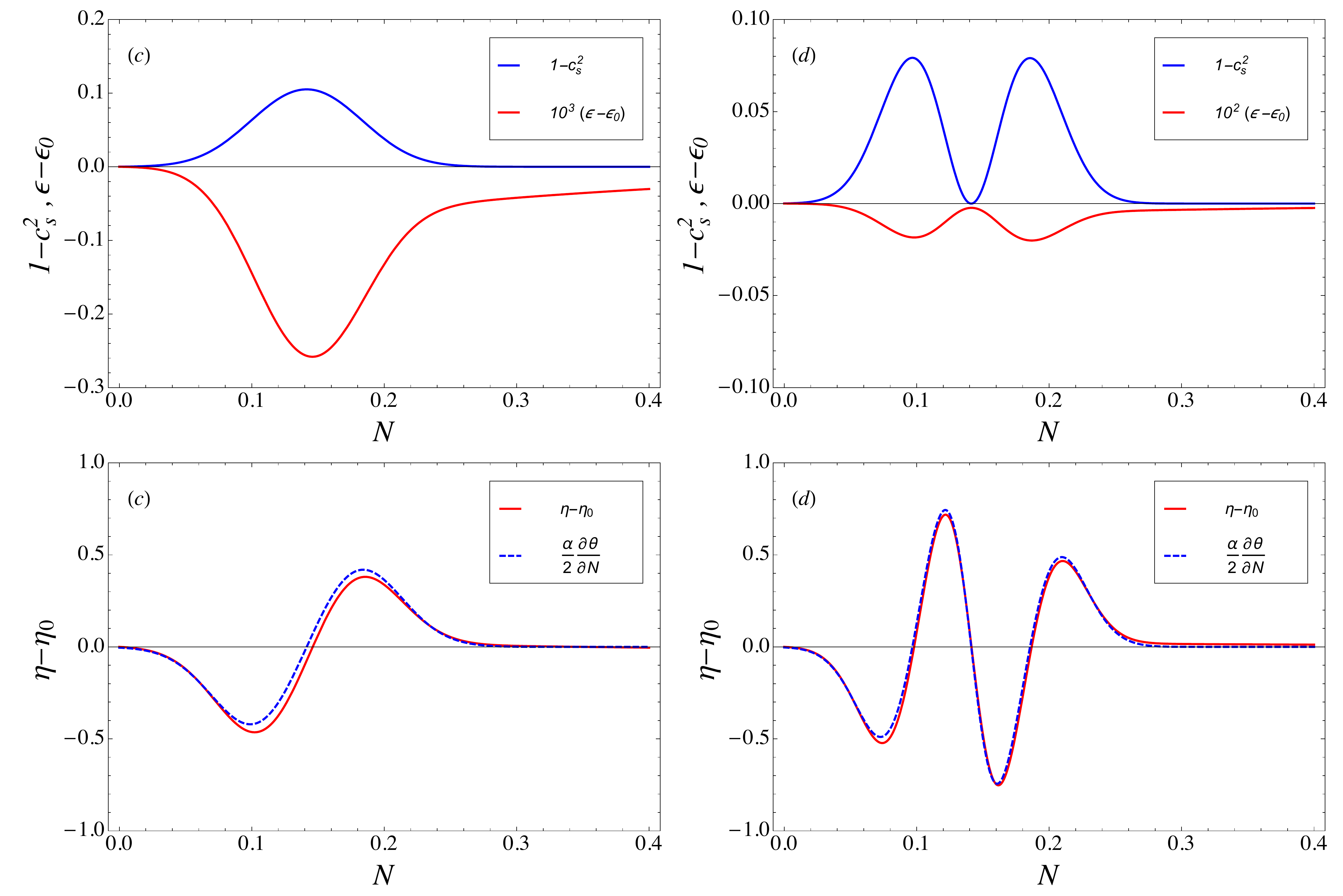} 
\end{center}
\caption{\footnotesize The figures show some relevant results concerning models $(c)$ and $(d)$ (left and right panels respectively). The top panels show $1 - c_s^2$ and $\Delta \epsilon = \epsilon - \epsilon_0$ as a function of $N$. The bottom panels show $\Delta \eta = \eta - \eta_0$ and $\alpha \partial_N \theta/2$  as a function of $N$. } \label{fig:plots-pofx}
\end{figure}

For our second example we will stay within the realm of $P(X, \phi)$-models, and consider a DBI theory~\cite{Silverstein:2003hf, Alishahiha:2004eh} parametrized by the following Lagrangian:\footnote{An alternative example, which renders similar results to those offered by this example, consists of $P(X, \phi) = X + g(\phi) X^2 - V(\phi)$, where $g(\phi)$ accomplishes the same role of $f(\phi)$.}
\be
P(X, \phi) = f^{-1}(\phi) \left[ 1 - \sqrt{ 1 - 2f(\phi) X } \right] - V(\phi). \label{Lagrangian-DBI}
\ee
Here, $f(\phi)$ corresponds to the warping factor describing a system of D3 branes on a warped background accomplished in certain string compactifications. $f(\phi)$ couples together the rapidity $\dot \phi$ and the vacuum expectation value $\phi$, allowing for departures from $c_s^2 = 1$ depending on the shape of $f(\phi)$. Observe that in order to ensure that we stay in the physical regime $1 - c_s^2 \geq 0$ at all times, we require $f(\phi) \geq 0$ everywhere. 
For definiteness, let us consider two choices for $f(\phi)$ characterized for being suppressed everywhere except for a region of finite support. Our first choice, model-$(c)$, consists of a Gaussian bump of the form
\be
f_{(c)}(\phi) = B \exp \left[ - \frac{(\phi - \phi_0)^2}{2 \Delta \phi^2} \right] ,
\ee
where $\phi_0$ and $\Delta \phi$ are the position and width of the bump, and $B$ is the amplitude modulating the coupling between $\phi$ and $X$. For our second choice, model-$(d)$, let us consider the following form for $f(\phi)$, consisting of a Gaussian bump multiplied by $(\phi - \phi_0)^2$:
\be
f_{(d)}(\phi) = B \, \frac{(\phi - \phi_0)^2}{\Delta \phi^2} \exp \left[ - \frac{(\phi - \phi_0)^2}{2 \Delta \phi^2} \right] .
\ee
Here again, $\phi_0$ and $\Delta \phi$ are the position and width of the bump, which this time consists of two peaks around $\phi_0$. These two alternatives for $f(\phi)$ induce a brief exchange between both kinetic and potential energy when the vacuum expectation value of $\phi$ hits the value $\phi_0$. Moreover, the non-trivial dependence of the Lagrangian on $X$ implies a sound speed different from unity during the duration of this exchange. To examine the effects of these sudden variations of the background, let us consider again a chaotic potential of the form
\be
V(\phi) = \frac{m^2}{2} \phi^2 ,
\ee
able to realize inflation within canonical single field slow roll inflation, with $60$ $e$-folds of inflation, and adjust the values of $\phi_0$, $\Delta \phi$, and $B$ to generate variations such that $\Delta \epsilon \ll \epsilon_0$ and $1 - c_s^2 \ll 1$. We have chosen (for both models)
\be
 \phi_0=15.387, \quad B= 5 \times 10^{9},  \quad m=5.875 \times 10^{-6}. \label{values-parameters-example2}
\ee
Moreover, in model-$(c)$ we set $\Delta \phi=0.005$ while in model-$(d)$ we have $\Delta \phi=0.004$. Figure \ref{fig:plots-pofx} shows the main numerical results for our two models. The top panels show the background solutions for both $\Delta \epsilon = \epsilon - \epsilon_0$ and $1 - c_s^2$ as a function of $e$-folds $N$. It may be seen that they happen in synchrony, as already anticipated by our general discussion linking both quantities. The bottom panels show both $\Delta \eta = \eta - \eta_0$ and $\alpha~ \partial_N \theta /2$ as functions of $N$. We find that there is a good agreement between both quantities for the values 
\be
\alpha_{(c)} \simeq - 0.53 , \qquad  \alpha_{(d)} \simeq - 0.52 .
\ee  
The resulting plots ratify the validity of our ansatz  (\ref{relation-eta-sound-2}). We have checked that this relation is ensured as long as the conditions $\Delta \epsilon \ll \epsilon_0$ and $1 - c_s^2 \ll 1$ are satisfied. As soon as these conditions are broken, one starts to observe a disagreement between both quantities. 

\subsection{Features in multi-field inflation}

Let us now take a look into the appearance of features within the framework of multi-field inflation. Concretely, we consider a multi-field model with a set of two scalar fields $\phi^a$, $a=1,2$, with a non-trivial sigma model metric $\gamma_{a b}$, and a Lagrangian given by
\be
\mathcal L =  - \frac{1}{2} \gamma_{a b} \partial \phi^a \partial \phi^b - V(\phi) , \label{L-multi-field}
\ee
where $V(\phi)$ is the scalar potential responsible for producing inflation. Given that we are focusing our interest on the case of multi-field models with only two fields, it is convenient to introduce a pair of unit vectors $T^a$ and $N^a$ tangent and normal to the inflationary trajectory $\phi^a = \phi^a (t)$, defined as
\bea
T^a &\equiv& \frac{1}{\dot \phi_0} \frac{d \phi^a}{d t} , \\ 
N_a &\equiv& \sqrt{\det \gamma} \epsilon_{a b} T^b, 
\eea
where $\epsilon_{a b}$ is the Levi-Civita symbol in two dimensions (notice that these definitions keep the orientation of the two vectors fixed). In addition, we have defined
\be
\dot \phi_0 \equiv \sqrt{\gamma_{b c} \dot \phi^b \dot \phi^c}.
\ee
As usual, the metric $\gamma_{ab}$ and its inverse $\gamma^{ab}$ are used to lower and rise indices on tensor living in the tangent space of the multi-field manifold parametrizing the model. The vectors $T^a$ and $N^a$ allow us to introduce the local rate of turn of the inflationary trajectory $\Omega$ defined to satisfy\footnote{Notice that this quantity has been denoted $\dot \theta$ in refs. \cite{Achucarro:2010da,Cespedes:2012hu} studying the effect of turns on the dynamics of curvature perturbations. Here we opt to use $\Omega$ in order to avoid any confusion with $\theta = 1 - c_s^2$.}
\be
\frac{D T^a}{d t} = - \Omega \, N^a ,
\ee
where $D T^a / dt = \dot T^a + \Gamma^{a}_{b c} T^b \dot \phi^c$ is the covariant time derivative of $T^a$ with $\Gamma^{a}_{b c}$ being the Christoffel symbols computed with the help of $\gamma_{ab}$. The quantity $\Omega$ is nothing but the angular velocity of the tangent vector $T^a$ as the inflationary trajectory bends. To continue, the equations of motion dictating the evolution of the fields $\phi^a$, deduced from the action (\ref{L-multi-field}), are given by:
\bea
\frac{D \dot \phi^a}{d t} + 3 H \dot \phi^a + V^a = 0, \label{eq-motion-multi}  \\ 
 3 H^2 = \frac{1}{2} \dot \phi_0^2 + V(\phi) . \label{Friedman-multi}
\eea
The two equations of motion for the scalar fields (\ref{eq-motion-multi}) may be projected along $T^a$ and $N^a$ respectively, giving us back
\bea
\ddot \phi_0 + 3 H \dot \phi_0 + V_\phi = 0 , \label{eqs-multi-field-1} \\  
\Omega = \frac{V_N}{ \dot \phi_0} ,  \label{eqs-multi-field-2}
\eea
where $V_\phi \equiv T^a V_a$ and $V_N \equiv N^a V_a$. On the other hand, by differentiating eq.~(\ref{Friedman-multi}) with respect to time, eq.~(\ref{eqs-multi-field-1}) allows us to show that:
\be
\epsilon = \frac{1}{2} \frac{\dot \phi_0^2}{H^2} . \label{epsilon-phi-dot}
\ee
A detailed analysis of the dynamics of the two scalar fluctuations in this class of theory shows that perturbations may be decomposed into curvature perturbations and a massive degree of freedom, associated to the fluctuations perpendicular to the inflationary trajectory~\cite{Gordon:2000hv, GrootNibbelink:2000vx, GrootNibbelink:2001qt}. The mass $m$ of the massive degree of freedom is found to be
\be
m^2 = V_{NN} + \epsilon  H^2 \mathbb{R} + 3 \Omega^2 ,
\ee
where $V_{NN} \equiv N^a N^b (V_{a b} - \Gamma_{a b}^{c} V_c)$, and $\mathbb{R}$ is the Ricci scalar computed out of $\gamma_{a b}$. Then, it turns out that if $m^2 \gg H^2$ it is possible to deduce a low energy effective theory for curvature perturbations alone, of the form (\ref{basic-S-def}), where the sound speed is given by~\cite{Achucarro:2010da}:
\be
c_s^{-2} = 1 + \frac{4 \Omega^2}{m^2 - 4 \Omega^2} .
\ee
In other words, the sound speed will always decrease as soon as the inflationary trajectory suffers a turn.

Now that we have a concrete relation giving us the sound speed in terms of quantities parametrizing the inflationary trajectory, we can understand the validity of relation~(\ref{relation-eta-sound-2}) under a new scope. First, it is important to appreciate that departures of the sound speed from the value $c_s = 1$ are exclusively due to turns of the inflationary trajectory $\Omega \neq 0$. In addition, eq~(\ref{eqs-multi-field-2}) tells us that each time there is a turn, the trajectory is pushed against the wall of the potential away from the minimum $V_N = 0$ (which, from a Newtonian point of view, it is a consequence of the angular momentum accompanying the turn). Then, each time there is a turn, the inflationary trajectory will tend to climb the wall of the potential, implying an increase in the value of the potential, and a consequential reduction of the kinetic energy $X = \dot \phi_0^2 / 2$. This has the cost of producing a small reduction of the value of $\epsilon$, as implied by eq.~(\ref{epsilon-phi-dot}). Therefore, a turn simultaneously produces both a reduction of the speed of sound $c_s^2 <1$ and a decrease of the value of $\epsilon$, finally leading to eq.~(\ref{relation-eta-sound}).

\subsubsection{Example 3: Localized turns in a two-field target space} \label{sec:example-multi-field}

\begin{figure}[t!]
\begin{center}
\includegraphics[width=1.0\textwidth,
]{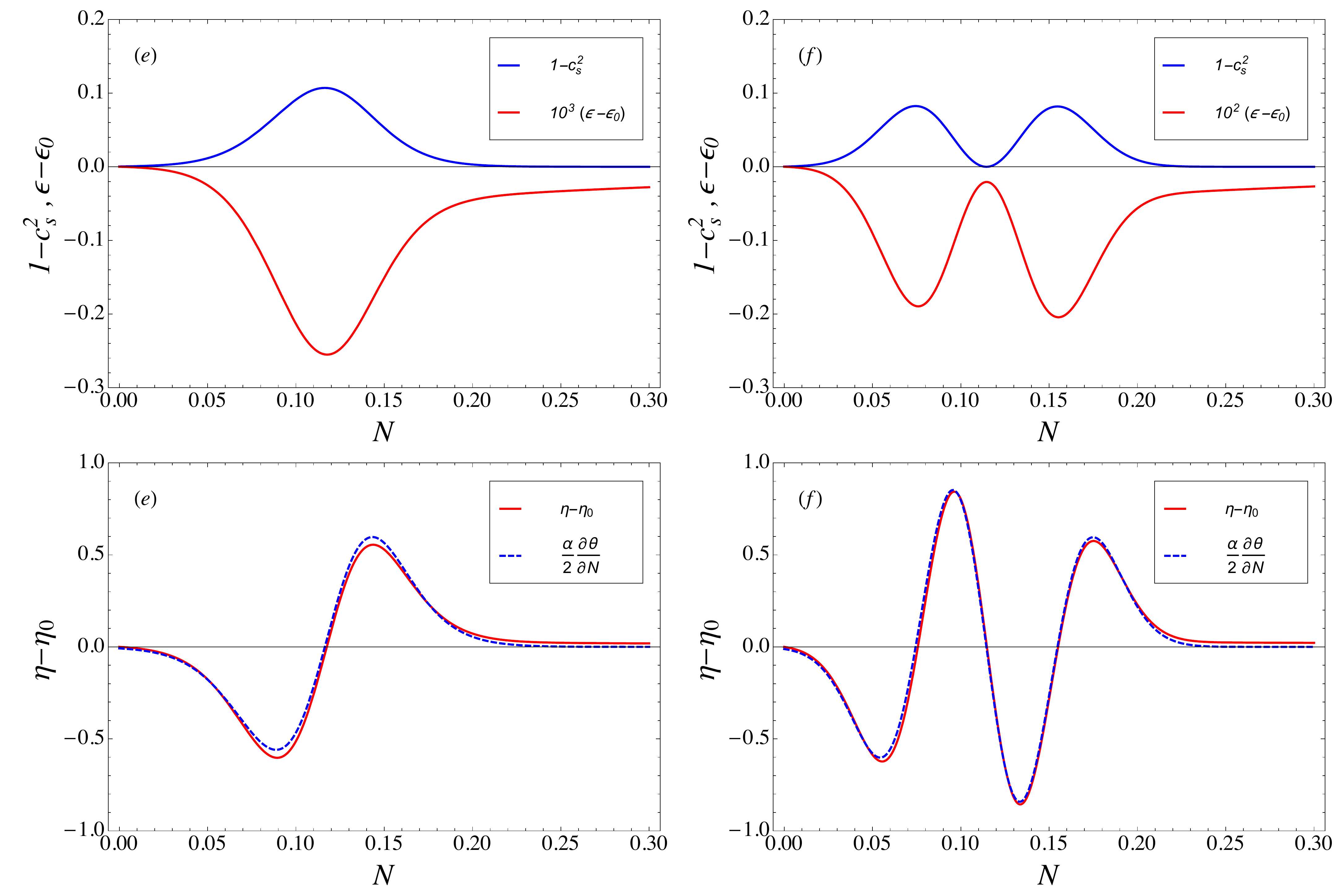} 
\end{center}
\caption{\footnotesize The figures show various relevant results related to the models $e$ and $f$ (left and right panels respectively). The top panels show $1 - c_s^2$ and $\Delta \epsilon = \epsilon - \epsilon_0$ as a function of $N$. The bottom panels show $\Delta \eta = \eta - \eta_0$ and $\alpha \partial_N \theta/2$  as a function of $N$.} \label{fig:plots-turns}
\end{figure}

We now proceed to examine a few concrete numerical examples of inflationary trajectories with turns, corroborating eq.~(\ref{relation-eta-sound}). Let us start by choosing the following notation for the two scalar fields:
\be
\phi^{a} = (\chi , \psi) .
\ee
We will consider a model where turns occur as a consequence of a feature appearing in the sigma model metric given $\gamma_{a b}$. To simplify matters, let us parametrize the metric in the following way\footnote{This type of model was already considered in ref.~\cite{Cespedes:2012hu} to study features in the power spectrum as the consequence of turns in multi-field models.}
\be
\gamma_{ab}=\left( \begin{array}{cc}
1 & f(\chi) \\
f(\chi) & 1+f^2(\chi) \end{array} \right), \label{sigma-model-metric}
\ee
where $f$ is a given function of the first field $\chi$ to be specified shortly. Notice that this metric has a unit determinant by construction. In addition, we choose a potential with the following generic form
\be
V(\chi,\psi)= v(\chi) +  \frac{M_{\psi}^2}{2} \psi^2, \label{multi-field-potential}
\ee
where $v(\chi)$ is a single field potential constructed to realize slow-roll inflation in the $\chi$-direction. On the other hand, the term with $M_{\psi}^2$ represents a mass term forcing $\psi$ to stabilize at the value $\psi = 0$. Notice that if $f(\chi) = 0$ the two fields would remain decoupled, and inflation would happen along the $\chi$-direction, as the $\chi$-field descends the slope of the potential $v(\chi)$, with $\psi$ stabilized at the value $\psi= 0$. More generally, if $f$ is a constant, then one can always redefine the fields and deduce that the inflationary trajectory consists of a straight path in the multi-field target space. To make things more concrete, let us consider a simple chaotic potential of the form
\be
v(\chi) = \frac{m^2}{2} \chi^2 ,
\ee
and choose initial conditions in such a way that inflation lasts for about $60$ $e$-folds. Next, let us consider two different choices for the profile of the function $f(\chi)$, characterized for being constant everywhere except for a region of finite support. For our first choice, model-$(e)$, let us consider the following function:
\be
f_{(e)}(\chi)= \frac{B}{2} \left( 1+ \tanh{\left[\frac{(\chi-\chi_0)}{\sqrt{2} \Delta\chi}\right]}  \right).  \label{G1}
\ee
This represents a step function, with a width $\Delta \chi$ centered at $\chi_0$ that changes from $f = 0$ to $f = B$. It is easy to see that this transition implies a single turn with non-vanishing rate of turn $\Omega$ inducing a non-trivial departure of $c_s^2$ away from unity. For our second choice, model-$(f)$, we consider a Gaussian function of the form 
\be
f_{(f)}(\chi)=  B   \exp \left[ - \frac{(\chi - \chi_0)^2}{2 \Delta \chi^2} \right] .  \label{G2}
\ee
This form of $f$ induces two consecutive turns in opposite directions. As a consequence, we obtain (for the same set of parameters) a non-trivial departure of $c_s^2$ away from unity in the form of two consecutive bumps. To solve the inflationary dynamics of these two models we consider the following values 
\be
B=2.5, \qquad M_\psi=  10^{-2}, \qquad m=5.89\cdot 10^{-6},Ê\qquad \Delta \chi = 0.005, \qquad  \chi_0 = 15.37 .
\ee
Figure~\ref{fig:plots-turns} shows the main numerical results for these two models. The top panels show the background solutions for both $\Delta \epsilon = \epsilon - \epsilon_0$ and $1 - c_s^2$ as a function of $e$-folds $N$. It may be seen that they happen in synchrony, as already anticipated by our general discussion linking both quantities. On the other hand, the bottom panels show both $\Delta \eta = \eta - \eta_0$ and $\alpha~ \partial_N \theta /2$ as functions of $N$. We find that there is a very good agreement between both quantities for the following values of $\alpha$:
\be
\alpha_{(e)} \simeq - 0.53 , \qquad  \alpha_{(f)} \simeq - 0.52 .
\ee  
Yet again, these numerical results emphasize the accuracy of our ansatz (\ref{relation-eta-sound-2}) to describe inflationary backgrounds with rapid deviations from quasi de Sitter.

\subsection{Accuracy of the power spectrum} \label{accuracy-power}

\begin{figure}[t!]
\begin{center}
\includegraphics[width=1.0\textwidth,
]{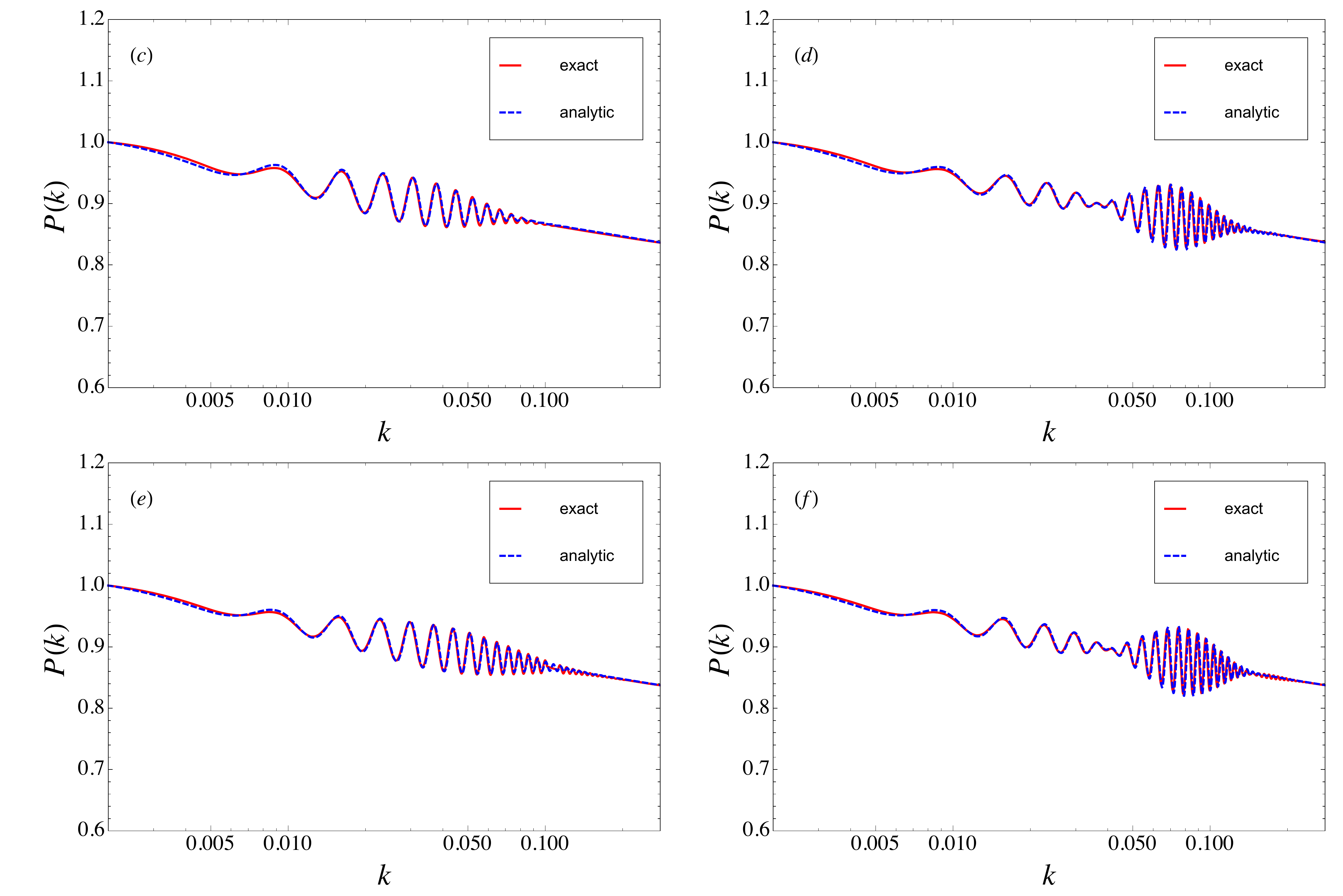} 
\end{center}
\caption{\footnotesize The figure shows a comparison between the power spectrum obtained by solving the mode functions of curvature perturbations exactly (continuous red curves) and our analytical approximation (dashed blue curves) using eq.~(\ref{psapp}).} \label{fig:power-spectrum-compared}
\end{figure}

Our previous examples (2 and 3) consisted of models that required values of $\alpha$ that were worryingly close to the value $\alpha = -1$. Indeed, let us recall that our expression for the power spectrum given by eq.~(\ref{psapp}) stops being valid for values of $\alpha$ close to $-1$, where other subleading terms start to dominate the final form of the power spectrum. Nevertheless, we find that for the aforementioned examples the power spectrum computed with~(\ref{eta-Power}) constitutes a very good approximation of the power spectrum computed directly by numerically solving the mode functions exactly. Figure~\ref{fig:power-spectrum-compared} compares the power spectrum obtained by solving the mode functions of curvature perturbations exactly (continuous red curves) and our analytical approximation (dashed blue curves) using eq.~(\ref{psapp}), for models $(c)$, $(d)$, $(e)$ and $(f)$ respectively. These plots confirm that the approximation offered by eq. (\ref{psapp}) gives an accurate representation of features in the power spectrum, in spite of the value $\alpha \sim - 0.5$.

The main reason for the good agreement between both power spectra is the sharpness of the features. Indeed, let us recall that eq.~(\ref{psapp}) is derived from (\ref{features-delta-2}) by keeping the leading terms according to the hierarchy (\ref{hierarchy-B-2}). This tells us that the power spectrum of eq.~(\ref{psapp}) is accurate as long as
\be
\frac{ 1+ \alpha }{  \Delta N }  \gg1 , \label{validity-power-spectrum}
\ee
where $\Delta N$ is the $e$-fold scale characterizing the rapidly varying background. If this condition is not fulfilled (for instance, because $\alpha \simeq -1$), then eq.~(\ref{features-delta-2}) tells us that the leading order term becomes:
\be
 \frac{\Delta \mathcal P}{\mathcal P_0 } (k) =  - \frac{5\alpha}{16k^3} \int_{-\infty}^{0} \!\!\! d \tau \,  \frac{\theta''' }{ \tau}   \, \sin (2 k \tau) \, . \label{features-delta-alpha=-1} 
\ee
Of course, this form of the power spectrum modifies our main result~(\ref{beta-alpha}), a situation that will not be examined in the present article.

\section{Resonant features} \label{sec:resonant-features}
\setcounter{equation}{0}

In this section we use our formalism to consider the case of resonant features. By resonant features we mean features in the power spectrum and bispectrum induced by an in\-fla\-tio\-na\-ry background with a certain periodic time-dependence~\cite{Flauger:2010ja, Behbahani:2011it, Behbahani:2012be, Adshead:2014sga}. We wish to show that our ansatz~(\ref{relation-eta-sound-2}) continues to constitute a good characterization of the time-dependence of resonant departures from quasi-de Sitter, and therefore our relation~(\ref{resu}) offers a concrete parametrization for the search of resonant features in the primordial spectra. Before con\-si\-de\-ring a concrete example, let us first study the implication of eq.~(\ref{relation-eta-sound-2}) analytically in the case of resonant features. To do so, let us consider a background characterized for having a varying sound speed with a harmonic dependence on the $e$-fold variable $N$:
\be
\theta (N) = \theta_0 \left[ 1 + \sin (\omega_0 (N - N_0) ) \right]  . \label{theta-resonant}
\ee
Here $\omega_0$ is the frequency of the oscillations (in $e$-fold units) and $\theta_0$ corresponds to the average value of $\theta (N)$. Equation~(\ref{relation-eta-sound-2}) then implies that $ \eta(N)$ is given by
\be
 \eta(N) = \eta_0 + \frac{ \alpha }{2} \,  \omega_0 \,  \theta_0 \cos (\omega_0 (N - N_0)  ) .
\ee
We may use these relations to obtain an expression describing the features in the power spectrum. Using $N = N_0 - \ln (\tau / \tau_0)$, which is valid for a constant expansion rate $H$, we find that (\ref{features-delta-2}) becomes
\bea
\frac{\Delta \mathcal P}{\mathcal P_0 } (k) &=& \frac{\theta_0}{2} + (1+\alpha) \theta_0  k \int_{-\infty}^{0} \!\!\! d \tau \,   \sin ( \omega_0 \ln ( \tau / \tau_0)  ) \, \sin (2 k \tau) , \label{power-resonant-1}
\eea
where the first term, $\theta_0/2$, comes from the non-oscillatory piece of eq.~(\ref{theta-resonant}). In fact, our discussion around eq.~(\ref{check-constant-speed}) tells us that we may absorb this piece in the definition of the featureless power spectrum $\mathcal P_0 \to \P_s =  \mathcal P_0 / c_0$, where $c_0  = \sqrt{1 - \theta_0}$. Then, integrating eq.~(\ref{power-resonant-1}) we finally obtain
\bea
\frac{\Delta \mathcal P}{\mathcal P_s } (k) &=&  (1+\alpha) \frac{\theta_0}{2}  \rho_0 \cosh (\pi \omega_0 / 2) \sin \Big( \omega_0 \ln ( 2 k |\tau_0| ) + \varphi_0 \Big)   , \label{power-resonant-analytic}
\eea
where $\rho_0$ and $\varphi_0$ are real functions of $\omega_0$ defined through the algebraic equation $\rho_0 e^{i \varphi_0} \equiv \Gamma(1 -  i \omega_0)$, where $\Gamma$ correspond to the usual $\Gamma$-function. This result may be more conveniently summarized as
\bea
\frac{\Delta \mathcal P}{\mathcal P_s } (k) &=&  A  \sin \left[ \omega_0 \ln  2 k +  \varphi \right]  ,
\eea
where $\varphi = \varphi_0 + \omega_0 \ln |\tau_0|$. This relation tells us that the power spectrum has a logarithmic dependence on the scale $k$. This result may be plugged back in eq.~(\ref{f_NL_features}) to obtain an expression for the bispectrum containing resonant features
\be
 f_{\rm NL}  \simeq  -  \beta_\alpha (k_1,k_2,k_3) A \omega_0^2  \sin \left[ \omega_0 \ln  (k_1 + k_2 + k_3) +  \varphi  \right]  , \label{f_NL_features-resonant}
\ee
where $\beta_\alpha$ is given by eq.~(\ref{resu}). This result may be compared with previous parametrizations used to search resonant features in the CMB data~\cite{Jackson:2013mka}, such as the one used by the Planck satellite to test resonant features in the non-Gaussian bispectrum~\cite{Ade:2015ava}, with the generic form given by
\be
B_{\rm res}(k_1,k_2,k_3)= g (k_1,k_2,k_3) \sin{\left[ C\ln{\left( k_1+k_2+k_3\right)}+\phi \right]} \label{plres}
\ee
where $g (k_1,k_2,k_3)$ is a function of the shape of the triangle configuration. In our case, the $\alpha$ parameter plays an important role in determining the shape of $g (k_1,k_2,k_3)$.

\subsection{Example 4: Resonant features in multi-field inflation}

\begin{figure}[t!]
\begin{center}
\includegraphics[width=1.0\textwidth,
]{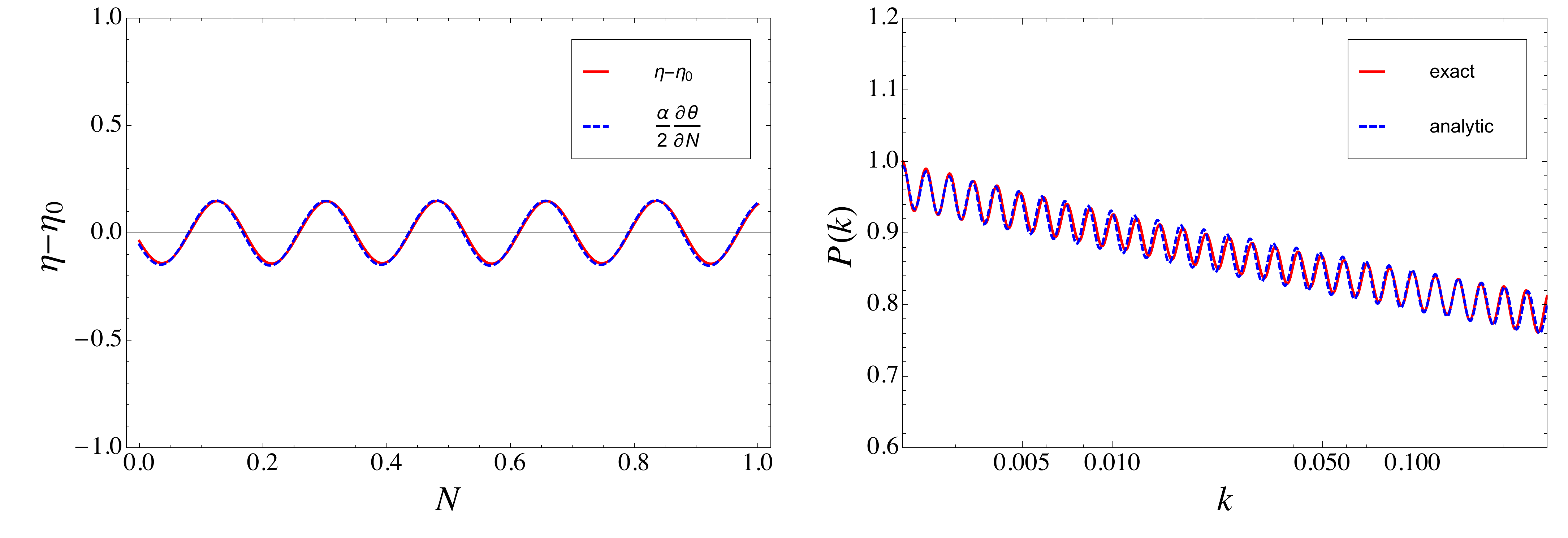} 
\end{center}
\caption{\footnotesize The left panel of the figure compares the background quantities $\eta - \eta_0$ and $\alpha \partial_N \theta /2$. The right panel compares the exact power spectrum computed numerically, with the analytical expression given by eq.~(\ref{power-resonant-analytic}).} \label{fig:resonant-plot}
\end{figure}

Let us now consider an example where the background offers a periodic dependence on time. To be concrete, we consider again the multi-field model of Section~\ref{sec:example-multi-field}, where the features are introduced via the function $f(\chi)$ appearing in the $\sigma$-model metric of eq.~(\ref{sigma-model-metric}). In this occasion we shall consider the following form for the function $f(\chi)$
\be
f(\chi) = B \cos \left( 2 \pi \frac{( \chi - \chi_0 )}{ \Delta \chi} \right) , \label{resonant-multi-field-f}
\ee
and keep the chaotic potential $v(\chi) = m^2 \phi^2 / 2$ for the inflationary sector $\chi$ within the multi-field potential (\ref{multi-field-potential}). The choice (\ref{resonant-multi-field-f}) ensures that the background quantities appearing in the EFT of curvature perturbations will inherit a periodic dependence on time. In particular, one finds that both $c_s$ and $\epsilon$ evolve periodically ---and in synchrony--- just as anticipated by eq.~(\ref{conjecture}), and one verifies the validity of eq.~(\ref{relation-eta-sound-2}). To offer an example, we have used the following values for the parameters of the present model:
\be
B = 1.5, \qquad  M_\psi=  10^{-2}, \qquad \Delta \chi = 0.05, \quad m=5.975 \times 10^{-6} .
\ee
These values ensure that the departures of $c_s^2$ and $\epsilon$ are small from their respective featureless background values $1$ and $\epsilon_0$. Figure~\ref{fig:resonant-plot} compares both $\eta - \eta_0$ and $\alpha \partial_N \theta /2$ for the value $\alpha = -0.53$, which is found to give the best match between the two functions.
Moreover, we find that~(\ref{theta-resonant}) constitutes a good fit of $\theta(N)$ with $\theta_0$ and $\omega_0$ given by:
\be
\theta_0 \simeq 0.016, \qquad \omega_0 \simeq 35.4.
\ee
With these values, it turns out that our expression for the resonant power spectrum of eq.~(\ref{power-resonant-analytic}) offers an accurate representation of the effects of the oscillatory time dependence of the background. The right panel of Figure~\ref{fig:resonant-plot} compares the power spectrum of the model computed numerically with the expression given by eq.~(\ref{power-resonant-analytic}), showing a very good agreement between both functions.

It is important to mention that we have also verified that eq.~(\ref{power-resonant-analytic}) offers an accurate expression for classes of models other than multi-field. Indeed, we have analyzed numerically the example of Section~\ref{example-DBI} for which we have chosen the warp factor $f(\phi)$ to have the same form of $f(\chi)$ in eq.~(\ref{resonant-multi-field-f}), and have found similar results to those presented here.

\section{Discussion and conclusions}  \label{sec:conclusions}
\setcounter{equation}{0}

While it is hard to conceive that observations could ever reveal the nature of the physics responsible for features, it is clear that, if they are present at all, both the power spectrum and bispectrum will contain complementary information about the source that generated them. Indeed, in this work we have seen that, under certain reasonable assumptions, it is possible to correlate features appearing in the bispectrum with those appearing in the power spectrum. Specifically, we have studied the generation of sharp features by adopting the perspective of effective field theory of inflation, where the primordial spectra of curvature perturbations are the consequence of the dynamics of a single scalar degree of freedom. In this framework, sharp features are the consequence of the simultaneous variations of the sound speed $c_s$ and the expansion rate, parametrized by $\epsilon$. 

The novelty of this work is that we have shown that two different kind of features, the ones caused by a perturbation in $\eps$ and the ones caused by a perturbation in $c_s$, are in fact not separate at all. After all, they follow from the same perturbed path that the inflaton takes through field space. In this way, in order to correlate the primordial power spectrum with the bispectrum, we had to uncover a universal relation linking $c_s$ and $\epsilon$, given by eq.~(\ref{newrel}). This relation ---valid for small but rapid variations of $c_s$ and $\epsilon$--- led to our main result, given by eq.~(\ref{f_NL_features}) supplemented with eq~(\ref{resu}). To show the validity of eq.~(\ref{newrel}), we offered several numerical examples of models allowing for sharp features.

The aim of this work therefore is to break the degeneracy between features in the power spectrum induced by perturbations in $\epsilon$, and features in the power spectrum induced by perturbations in $c_s$. Looking at the power spectrum alone, they cannot be told apart. With this work in hand one can, however, link the features in the power spectrum to those appearing in the bispectrum. One can extract a value for $\alpha$, which indicates exactly to what extent the feature is caused by a perturbation in $\epsilon$, and to what extent by a perturbation in $c_s$.

Our results leave several questions unanswered that deserve to be addressed. Let us mention a few possible directions that we find interesting: 
\begin{enumerate}

\item One may consider the general problem of using the present methods to study the appearance of features in higher order $n$-point correlation functions. The present results suggest that features in higher $n$-point correlation functions may be expressed in terms of features in the power spectrum. More concretely, it would already be interesting to learn how the tri-spectrum depends on features appearing in the power spectrum.

\item As emphasized in Section~\ref{accuracy-power}, our main result (\ref{f_NL_features}) stops being valid for values of $\alpha$ violating condition~(\ref{validity-power-spectrum}), with $\alpha$ the proportionality parameter $\alpha$ between $\eta$ and $d\theta / dN$ . Thus, a direction for further research could be to identify and analyze models in which  $\alpha$ is sufficiently close to $-1$. In this case, we need to reconsider the relation between the perturbations in the power spectrum and $\eta$ and $\theta$, as terms which are subleading for $\alpha\neq 1$ become dominant. 

\item In this work, $\alpha$ is just a constant that has been adjusted to satisfy eq.~(\ref{relation-eta-sound}) in each model. Nevertheless, it is clear that $\alpha$ is a model dependent quantity. Thus, it might be possible to deduce analytical expressions for $\alpha$ in terms of parameters characterizing the models hereby examined (\emph{i.e.} $P(X,\phi)$ and multi-field models of inflation). 

\item We have not considered models in which the slowly varying part of the sound speed $c_0$ remains far from $1$. Our first  numerical example (Section~\ref{sec:example-1}) shows that a background sound speed different from $1$ does not spoil the relation between $\eta$ and $d\theta / dN$. However, in this situation there are new operators in the cubic action (\ref{final-S3-int}) for $\R$ inducing features in the bispectrum, not taken into account in the final result~(\ref{f_NL_features}). 

\item To correlate the features in the spectra we have assumed a standard single field effective theory (\ref{R-2-action}) describing the dynamics of curvature perturbations, in which the sound speed is the main parameter describing departures from canonical single field inflation. However, as already noticed in~\cite{Cheung:2007st}, there are more general classes of effective field theories where the dispersion relation of curvature perturbations contains non-trivial dependences on the scale~\cite{Gwyn:2012mw, Gwyn:2014doa}. Therefore, it would be interesting to analyze the way in which the primordial spectra are correlated within other classes of effective field theories.

\item We have focussed our efforts exclusively on the case of sharp features, where the hie\-rar\-chy of eq.~(\ref{hierarchy-B-1}) was assumed to characterize every rapidly varying background quantity. We see no obstructions towards obtaining a more general expression correlating features in the bispectrum with those of the power spectrum in the case where the variation of background quantities is not so restricted (as in the case of ref.~\cite{Achucarro:2012fd}, for the particular case of features emerging exclusively from a varying sound speed.). 
\end{enumerate}
We leave these interesting challenges as open problems.

To conclude, if some of the not-so-significant-yet features that Planck is seeing now turn out to be real, our work provides a framework to analyze them. If not, it can be equally useful in order to confirm the paradigm of plain, featureless single field slow roll inflation.

\subsection*{Acknowledgements}

We would like to thank Ana Ach\'ucarro, Vicente Atal, Jinn-Ouk Gong and Spyros Sypsas for helpful discussions and insightful comments on this manuscript. This work was supported by the Fondecyt project number 1130777 (GAP), by the ``Anillo'' project ACT1122 funded by the ``Programa de Investigaci\'on Asociativa" (GAP \& SM \& GP), by the Fondecyt 2015 Postdoctoral Grant 3150126 (SM), and by the CONICYT-PCHA/MagisterNacional/2013-221320624 (AS).

\begin{appendix} 

\renewcommand{\theequation}{\Alph{section}.\arabic{equation}}
\section{Perturbation theory for features} \label{app1}
\setcounter{equation}{0}

In this appendix we summarize the in-in formalism of perturbation theory in quasi-de Sitter space-times~\cite{Maldacena:2002vr, Weinberg:2005vy} that we use to compute the power spectrum and bispectrum with features. More details may be found in ref.~\cite{Palma:2014hra} and references therein. 

\subsection{The in-in formalism for features}

As a starting point, let us consider the action~(\ref{basic-S-def}), together with eqs.~(\ref{R-2-action}) and~(\ref{S-R-3-second}), describing the dynamics of curvature perturbations $\R$ in comoving gauge. Then, it is useful to introduce the canonically normalized curvature perturbation $u(t,\x)$ through the following field reparametrization:
\be
u = z  \, \mathcal{R}, \qquad z = \sqrt{2 \epsilon} \frac{a}{c_s}. \label{u-R}
\ee
Doing this change of variable allows us to split the quadratic part of action~(\ref{basic-S-def}) as $S^{(2)} = S_0^{(2)} + S_{\rm int}^{(2)}$, where the zeroth order part $S_0^{(2)}$ and the interacting parts $S_{\rm int}^{(2)}$, are respectively given by:
\bea
S_0^{(2)}&=&  \frac{1}{2}\int d^3x~d\tau\left[u'^2 - (\nabla u)^2+\frac{2}{\tau^2}u^2 \right] , \label{S-0-2} \\
S_{\rm int}^{(2)}&=&  \frac{1}{2}\int d^3x \, d \tau  \left[   \theta ( \tau )  (\nabla u)^2 + \frac{1}{\tau^2} \delta(t) u^2 \right] , \label{S-int-2}
\eea
where $\delta(t)$ is a rapidly varying background quantity, given by:
\be
\delta = \delta_{H} -  \tau \theta'  + \frac{\tau^2}{2} \theta'' ,Ê\qquad \delta_H = -  \frac{1}{2} \tau \eta' . Ê\label{complete-delta}
\ee
Notice that the zeroth order piece, $S_{0}^{(2)}$, corresponds to the conventional quadratic action describing a scalar perturbation in a de Sitter space-time sliced with the help of cosmological coordinates.  On the other hand, $S_{\rm int}^{(2)}$ contains coefficients induced by the rapidly varying parts of the background. Of course, we also have the cubic contribution to the total action, which, to leading order in the rapid background quantities, is given by~(\ref{final-S3-int}). This means that, with the purpose of employing the in-in formalism, we may finally split the entire theory as 
\be
S = S_0^{(2)} + S_{\rm int},
\ee
where $S_{\rm int}$ is the interacting part of the action, up to cubic order, and given by
\be
S_{\rm int} = S^{(2)}_{\rm int} + S^{(3)}_{\rm int} , \label{interaction-action}
\ee
where $S^{(2)}_{\rm int}$ is given by eq.~(\ref{S-int-2}) and $S^{(3)}_{\rm int}$ is given by eq.~(\ref{final-S3-int}). Equation~(\ref{S-0-2}) informs us that the $u$-field corresponds to the canonically normalized adiabatic curvature perturbation with a canonical momentum given by $\pi = \partial \mathcal L / \partial u'$ which, to leading order, reduces to $\pi = u'$. Both $u$ and $\pi$ satisfy the equal-time canonical commutation relation
\be
\left[ u(\x,\tau) , \pi (\y, \tau) \right] = i \delta(\x - \y) ,\label{quant-commut-rel}
\ee
where $\delta(\x - \y)$ is the Dirac delta function. The field $u(x,\tau)$ and its canonical momentum $\pi (x,\tau)$, satisfying this commutation relation, may be written in terms of the interaction-picture field $u_I(x,\tau)$ evolved in time with the help of the propagator $U(\tau)$ as:
\bea
&& u(\x,\tau) = U^{\dag}(\tau) \,  u_I(\x,\tau) \, U(\tau) , \label{u-complete} \\
&& \pi (\x,\tau) = U^{\dag}(\tau) \,  u_I' (\x,\tau) \, U(\tau) . \label{pi-complete}
\eea
Here the interaction-picture field $u_I(x,\tau)$ has the form of a free field, which may be expanded in Fourier modes as
\be
u_I (\x,\tau) = \frac{1}{(2 \pi)^{3}} \int d^3 k \left[ a_\k u_k(\tau) e^{i \k \cdot \x}  + a_\k^{\dag} u_k^{*}(\tau) e^{-i \k \cdot \x}   \right] , \label{u-I-pict}
\ee
where $a_\k^{\dag}$ and $a_\k$ are creation and annihilation operators satisfying the usual commutation relations for particle states in a Fock space:
\be
\big[ a_\k , a_\p^{\dag} \big] = (2\pi)^3 \delta(\k - \p). 
\ee
On the other hand, $u_k(\tau)$ represents the normalized solution to the linear equation of motion in momentum space derived from~(\ref{S-0-2}):
\be
u_{k}'' + \left(  k^2  - \frac{2}{\tau^2} \right) u_{k} = 0. \label{eq-of-motion}
\ee
The solution to this equation is obtained by selecting the Bunch-Davis vacuum, and is given by:
\be
u_k(\tau) = \frac{1}{\sqrt{2 k}} \left( 1 - \frac{i}{k \tau} \right) e^{- i k \tau} . \label{BD-solution}
\ee
Returning to~(\ref{u-complete}) and~(\ref{pi-complete}), the propagator is given by
\be
U(\tau) = \mathcal T \exp \left\{ - i \int^{\tau}_{-\infty_+} \!\!\!\!\!\! d\tau' H_I (\tau') \right\} ,
\ee
where $\mathcal T$ stands for the standard time ordering symbol and $\infty_+ = (1 + i \epsilon) \infty$ is the prescription isolating the in-vacuum in the infinite past. Furthermore, $H_I$ is the interaction picture Hamiltonian given by
\be
H_I (\tau) = U^\dag(\tau)  H_{\rm int}  U(\tau) , \label{def-int-pict-ham}
\ee
where $H_{\rm int}$ is the interaction Hamiltonian derived from~(\ref{interaction-action}). In particular, it is possible to show that the quadratic contribution to the interaction-picture Hamiltonian has the form:
\be
H_{I}^{(2)} (\tau) = - \frac{1}{2} \int d^3x  \left[   \theta ( \tau )  (\nabla u_I)^2 + \frac{\delta (\tau) }{\tau^2} u_I^2 \right] . \label{int-pict-ham-2}
\ee
On the other hand, the cubic contribution to the interaction picture hamiltonian, $H_I^{(3)}$, is found to be given by
\be
H_I^{(3)}=m_{Pl}^2\int d^3x~a^3\epsilon\left[(3\theta+\eta) \R_I\dot{\R}_I^2 + \frac{1}{a^2} ( \tau\theta' - \eta) \R_I(\nabla\R_I)^2\right].  \label{HI3}
\ee
Notice that we have opted to express $H_I^{(3)}$ in terms of $\R_I = - u_I H_0 \tau/ \sqrt{2 \epsilon_0}$ rather than $u_I$. These expressions allow us to compute the two- and three-point correlation functions for $u$ perturbatively. To first order in the rapid background quantities, one finds:
\bea
 \langle u(\mathbf{x},\tau)u(\mathbf{y},\tau)\rangle = \langle 0| u_I(\mathbf{x},\tau)u_I(\mathbf{y},\tau)|0\rangle+i\int_{-\infty}^{\tau}d\tau'\langle 0|[H_I(\tau'),u_I(\mathbf{x},\tau')u_I(\mathbf{y},\tau')]|0\rangle,  \qquad   \label{2p}  \\
 \langle u(\mathbf{x},\tau)u(\mathbf{y},\tau)u(\mathbf{z},\tau)\rangle  =  i\int_{-\infty}^{\tau}d\tau'\langle 0|[H_I^{(3)}(\tau'),u_I(\mathbf{x},\tau')u_I(\mathbf{y},\tau')u_I(\mathbf{z},\tau')]|0\rangle. \qquad \label{3p}
\eea
These expressions conform the basis to compute the power spectrum and bispectrum with features, which is done in what follows.

\subsection{Primordial spectra with features} \label{derivation-P-B}

In order to define the power spectrum, and later on the bispectrum, we write the adiabatic perturbations in Fourier space by introducing the mode function $\hat \R_{\k} (\tau) $ as:
\be
\R (\x,\tau) = \frac{1}{(2 \pi)^{3}} \int d^3 k \, \hat \R_{\k} (\tau) e^{i \k \cdot \x} .
\ee
Then, the dimensionless power spectrum $\mathcal P_{\mathcal R} ( k , \tau)$, evaluated at a given time $\tau$, is defined via the equation:
\be
\langle \hat \R_{\k} (\tau) \hat \R_{\p} (\tau) \rangle = (2 \pi )^3 \delta (\k + \p) \frac{2 \pi^2 }{k^3} \mathcal P_{\mathcal R} ( k , \tau)  . \label{power-def}
\ee
Putting together (\ref{u-R}) and (\ref{power-def}) we deduce an expression for the power spectrum $\mathcal P_{\mathcal R} ( k , \tau)$ in terms of the two-point correlation function of the canonically normalized $u$-field
\be
\mathcal P_{\mathcal R} ( k , \tau) =  \frac{k^3}{2 \pi^2  z^2}   \int_x \langle u(\x,\tau) u(0,\tau)  \rangle  e^{- i \k \cdot  \x} ,\label{power-spectrum-def}
\ee
where $\int_x$ stands for $\int d^3 x$. Notice that one of the fields has been conveniently evaluated at $\y = 0$, allowed by the homogeneity and isotropy of the background. Similarly, the bispectrum $B_{\R }(\k_1 , \k_2 , \k_3, \tau)$ is conventionally defined as:
\be
 \langle \hat \R_{\k_1} (\tau) \hat \R_{\k_2} (\tau) \hat \R_{\k_3} (\tau) \rangle = (2 \pi)^3 \delta (\k_1 + \k_2 + \k_3 ) B_\R (\k_1 , \k_2 , \k_3, \tau) .
\ee
This expression may be inverted to give the bispectrum in terms of the three-point correlation function of the $u$-field as
\be
B_\R (\k_1 , \k_2 ,\k_3 , \tau) = \frac{1}{z^3}  \int_x  \int_y  \langle u (\x,\tau) u (\y,\tau) u (0,\tau) \rangle e^{- i \k_1 \cdot \x - i \k_2 \cdot \y} , \label{bispectrum-def}
\ee
where $\k_3 = - \k_2 - \k_1$. Just as we did with eq.~(\ref{power-spectrum-def}), we have evaluated one of the fields at the comoving coordinate $\z = 0$. Finally, let us recall that our goal is to compute correlation functions in the long wavelength limit $| \tau k | \ll 1$, which gives us the spectra at the end of inflation. Thus, the power spectrum and bispectrum we are interested in correspond to the following formal limits:
\bea
\P_{\R} ( k) &\equiv& \lim_{|\tau| \to 0} \P_{\R} ( k , \tau) , \\
B_\R (\k_1 , \k_2 ,\k_3 ) &\equiv& \lim_{|\tau| \to 0} B_\R (\k_1 , \k_2 ,\k_3 , \tau) .
\eea
Given that $\R$ is constant after horizon crossing, these expressions give us the initial conditions for adiabatic perturbations outside the horizon for the hot Big-Bang era, in terms of background parameters at horizon crossing time, during inflation.

To compute the primordial power spectrum with features we just need to plug (\ref{2p}) into (\ref{power-spectrum-def}). This leads to the following form of the power spectrum
\be
\P=\P_0 +\Delta \P , \label{complete-power-spectrum-appendix}
\ee
where $\P_0$ corresponds to the featureless power spectrum given by:
\bea
\mathcal P_0 (k) = \lim_{\tau \to 0} \frac{k^3}{2 \pi^2 z^2} | u_k(\tau) |^2  = \frac{H_0^2 }{8  \pi^2 \epsilon_0}  .  \label{zeroth-power}
\eea
In fact, if we had considered slow-roll corrections to the zeroth order mode, we would have obtained the more accurate expression (used in the present work)
\bea
\mathcal P_0 (k) =  \frac{H_0^2}{8\pi^2\epsilon_0} \left( \frac{k}{k_*} \right)^{n_s - 1},   \label{zeroth-power-v2}
\eea
where $n_s$ is the spectral index parametrizing the small scale dependence of the power spectrum, and $k_*$ is a pivot scale fixed by mode that exited the horizon when the background was characterized by an expansion rate $H_0$.
To continue with eq.~(\ref{complete-power-spectrum-appendix}), $\Delta \P$ is the part containing the features, found to be given by
\bea
\frac{\Delta \mathcal P}{\mathcal P_0 } (k) &=& \lim_{k |\tau| \to 0}   i ( 2 c_0^3 k^3 \tau^2 )  \int^{\tau}_{-\infty} \!\!\!\!\!\! d\tau' \, \int_x  e^{- i \k \cdot  \x}   \langle 0|  \left[ H_I(\tau') ,u_I(\x,\tau) u_I(0,\tau) \right] | 0 \rangle \nn\\
&=&  k \int_{-\infty}^{0} \!\!\! d \tau \, \left[  - \theta + \frac{\delta_H}{k^2\tau^2}  + \frac{2 \delta_H}{k^4\tau^4}  - \frac{1}{k^4\tau^3} \frac{d \delta_H}{d \tau} \right] \, \sin (2 k \tau) , \label{dp}
\eea
where $\int_x$ stands for $\int d^3 x$. Finally, we may compute the bispectrum in a similar way by plugging (\ref{3p}) into (\ref{bispectrum-def}). This step gives us
\bea
\Delta B (\k_1, \k_2, \k_3) \! = \!   \frac{2 \epsilon_0}{i H_0^2} \R_1(0) \R_2 (0) \R_3(0)  \! \int_{- \infty}^{0} \!\!\!\!\!\! d \tau \Bigl( \frac{3\theta+\eta}{\tau^2} \left[ \R_1 (\tau) \R_2' (\tau) \R_3' (\tau) + sym \right]^* + {\rm c.c.} \nn\\
 -  \frac{\tau\theta' - \eta}{\tau^2}   \left[ \k_2 \cdot \k_3 \,  \R_1 (\tau) \R_2 (\tau) \R_3 (\tau) + sym \right]^* + {\rm c.c.} \Bigr), \qquad \qquad \qquad 
\eea
where $\R_i (\tau) \equiv \R (\k_i, \tau)$ is the wave function for comoving curvature perturbations obtained from~(\ref{BD-solution}):
\be
 \R_k (\tau) = i \frac{H_0}{2 \sqrt{\epsilon_0 k^3 }} \left(1 +  i k \tau \right) e^{- i k \tau} . \label{R-I}
\ee
The previous relation for the bispectrum may be further simplified. On the one hand, notice that 
\be
\k_1 \cdot \k_2 + \k_2 \cdot \k_3 +\k_3 \cdot \k_1  =  - \frac{1}{2} ( k_1^2 + k_2^2 + k_3^2) ,
\ee
which is valid as long as $\k_1 + \k_2 + \k_3 = 0$. Then, by using eq.~(\ref{R-I}) and assuming that the functions $\theta$ and $\eta$ are odd functions of conformal time $\tau$. Then we get
\bea
\Delta B (\k_1, \k_2, \k_3)   \! &=& \!   
  \frac{2 \pi^4 \P_0^2}{(k_1 k_2 k_3)^{3}}\int_{-\infty}^\infty d \tau ~i ~e^{i K \tau} \times\nn\\
&& ~ \Bigl[  (3\theta+\eta)    \left[ -(k_2k_3)^2-(k_3k_1)^2-(k_1k_2)^2 + i k_1 k_2 k_3 \tau (k_2k_3+k_3k_1+k_1k_2)   \right]         \nn\\
&& ~  +  \frac{\eta-\tau\theta' }{2\tau^2} \left[k_1^2+k_2^2+k_3^2  \right] (1-ik_1\tau) (1-ik_2\tau) (1-ik_3\tau)             \Bigr], \label{Delta-B}
\eea
where $K = k_1 + k_2 + k_3$, and where we have identified  $\P_0 = H_0^2 / 8  \pi^2 \epsilon_0 $. These are the results needed to compute the final expressions linking features in the bispectrum with those of the power spectrum.

\end{appendix}

\end{document}